\documentclass[a4paper,11pt]{article}
\pdfoutput=1

\usepackage{jcappub}
\usepackage{physics}

\newcommand*{\beqa}{\begin{eqnarray}}
\newcommand*{\eeqa}{\end{eqnarray}}

\newcommand*{\p}{\partial}

\newcommand*{\phid}{\dot{\phi}}

\newcommand*{\rhod}{\dot{\rho}}

\newcommand*{\gtilde}{\widetilde{g}}
\newcommand*{\Rtilde}{\widetilde{R}}
\newcommand*{\Ttilde}{\widetilde{T}}

\newcommand*{\Xtilde}{\widetilde{X}}

\newcommand*{\nablatilde}{\widetilde{\nabla}}
\newcommand*{\tilnab}{\widetilde{\nabla}}
\newcommand*{\tilbox}{\widetilde{\Box}}

\newcommand{\tila}{\widetilde{a}}
\newcommand{\tilc}{\tilde{c}}
\newcommand{\tild}{\widetilde{d}}
\newcommand{\tilf}{\widetilde{f}}
\newcommand{\tilg}{\widetilde{g}}
\newcommand{\tilh}{\widetilde{h}}
\newcommand{\tilk}{\widetilde{k}}

\newcommand{\tiln}{\widetilde{n}}
\newcommand{\tilp}{\widetilde{p}}
\newcommand{\tilr}{\widetilde{r}}
\newcommand{\tils}{\widetilde{s}}
\newcommand{\tilt}{\widetilde{t}}
\newcommand{\tilu}{\widetilde{u}}
\newcommand{\tilw}{\widetilde{w}}
\newcommand{\tilx}{\widetilde{x}}
\newcommand{\tilz}{\widetilde{z}}

\newcommand*{\tilC}{\widetilde{C}}
\newcommand*{\tilD}{\widetilde{D}}
\newcommand*{\tilF}{\widetilde{F}}
\newcommand*{\tilG}{\widetilde{G}}
\newcommand*{\tilH}{\widetilde{H}}
\newcommand*{\tilI}{\tilde{I}}
\newcommand*{\tilL}{\widetilde{L}}
\newcommand*{\tilN}{\widetilde{N}}
\newcommand*{\tilP}{\widetilde{P}}
\newcommand*{\tilR}{\widetilde{R}}
\newcommand*{\tilS}{\widetilde{S}}
\newcommand*{\tilT}{\widetilde{T}}

\newcommand{\tillambda}{\widetilde{\lambda}}

\newcommand{\tilrho}{\widetilde{\rho}}
\newcommand{\tiltau}{\widetilde{\tau}}
\newcommand{\tilomega}{\widetilde{\omega}}

\newcommand*{\tilGamma}{\widetilde{\Gamma}}
\newcommand*{\tilPhi}{\widetilde{\Phi}}

\newcommand*{\A}{{\cal A}}
\newcommand*{\B}{{\cal B}}
\newcommand*{\I}{{\cal I}}

\title{Disformal invariance of cosmological observables}

\author[a]{Takeshi Chiba,}
\author[b,1]{Fabio Chibana~\note{Corresponding author.}}
\author[b]{and Masahide Yamaguchi}

\affiliation[a]{Department of Physics, College of Humanities and Sciences, Nihon University, \\
                Tokyo 156-8550, Japan}
\affiliation[b]{Department of Physics, Tokyo Institute of Technology, \\
			    Tokyo 152-8551, Japan}

\emailAdd{chiba@phys.chs.nihon-u.ac.jp}
\emailAdd{chibana@th.phys.titech.ac.jp}
\emailAdd{gucci@phys.titech.ac.jp}

\abstract{
We study the frame dependence/independence of cosmological observables under disformal transformations, extending the previous results regarding conformal transformations, and provide the correspondence between Jordan-frame and Einstein-frame variables.
We consider quantities such as the
gravitational constant in the Newtonian limit, redshift, luminosity and angular diameter distances, as well as the distance-duality relation. 
Also,  the Boltzmann equation, the observed specific intensity, and 
the adiabaticity condition are discussed. 
Since the electromagnetic action changes under disformal transformations, photons in the Einstein frame no longer propagate along null geodesics. As a result, several quantities of cosmological interest are modified. 
Nevertheless, we show that the redshift is invariant and the distance-duality relation (the relation between the luminosity distance and the angular diameter distance)  
still holds in general spacetimes even though the reciprocity relation (the relation between two geometrical distances) is modified. 
}

\arxivnumber{2003.10633}

\begin{document}
\maketitle
\flushbottom

\section{Introduction}

The extension of gravity theory is now paid particular attention to. 
Our Universe has experienced a phase of accelerated expansion not 
only in the early stages of its evolution but also in the late 
Universe.
A scalar degree of freedom might be responsible for these accelerated 
expansions, which motivates us to extend {general relativity} 
to the so-called scalar-tensor theories of gravity. 
Such scalar-tensor theories of gravity~\cite{Fierz:1956zz,Jordan:1959eg,Brans:1961sx,Bergmann:1968ve,Nordtvedt:1970uv,Wagoner:1970vr} 
are usually formulated in the so-called Jordan frame, where the 
metric tensor is minimally coupled to the matter sector.  
On the other hand, when we perform a conformal transformation to the 
metric, the action can be reduced to that of the Einstein gravity, 
but the metric is now non-minimally coupled to the matter sector~\cite{dicke}.  
Therefore, an important question, which has been heavily 
discussed, is ``which frame actually describes physics?''.

In a previous paper~\cite{cy}, two of us (see also refs.~\cite{ds,catena})
studied this problem and found that the cosmological observables/
relations (redshift, luminosity distance, and temperature 
anisotropies) are frame-independent under the conformal  
transformation. 
Moreover, the frame dependence/independence of the adiabaticity 
condition and curvature perturbations was investigated in refs.~\cite{cy,Gong:2011qe}. 
Such frame dependence was also discussed at quantum level through 
one-loop corrections~\cite{Kamenshchik:2014waa,Ruf:2017xon,Ohta:2017trn}.

However, the most general metric transformation involving a scalar 
field that respects causality and the weak equivalence principle is 
the so-called disformal transformation~\cite{bekenstein}.
Since the causal structure changes under a disformal transformation, 
the propagation of photons is modified.
Hence, it is not clear whether cosmological observables are still 
frame-independent.
Several previous studies have argued that some cosmological 
observables --- such as the distance-duality relation and the 
spectrum of the cosmic microwave background (CMB) --- are frame-dependent. 
However, since the disformal transformation, as in the case of 
conformal ones, is merely a change of variables or units, we believe 
that the physical laws and any observable quantity should not depend 
on a particular set of variables, being disformal-frame independent. 
In fact, the invariance of physical quantities such as cosmological 
perturbations, causal structure, propagation speed, and number of 
degrees of freedom was already shown~\cite{Minamitsuji:2014waa,Motohashi:2015pra,Watanabe:2015uqa,Domenech:2015hka,Domenech:2015tca}, and some of our results in this paper overlap with those presented in these references.

In this paper, after presenting the relations between the variables 
in the Jordan and Einstein frames, we calculate, in both frames, the 
equations and observables in classical gravitational theory, 
including the geodesic equations and the effective gravitational 
constant~(section~\ref{sec:disformal_transformations_and_geodesics}); 
the redshift, luminosity distance, angular diameter distance, and the 
distance-duality relation~(section~\ref{sec:cosmology}); 
the photon distribution function and adiabaticity condition~(section~\ref{sec:perturbed_observables}).
At each step, we check if the quantities are frame-independent or 
frame-dependent. 
In appendix~\ref{sec:emt}, several relations of the matter energy-momentum tensor are given. 
Throughout this work, we use units such that $c=\hbar=1$, unless 
otherwise stated.

\section{Disformal transformation and geodesics}
\label{sec:disformal_transformations_and_geodesics}

\subsection{Disformal transformation}

The disformal transformation between  Jordan frame metric, 
$\gtilde_{\mu \nu}$, and the Einstein frame metric, 
$g_{\mu \nu}$, is defined as~\cite{bekenstein}
\begin{equation} \label{eq:metric_transf}
	\gtilde_{\mu \nu} = \A(\phi,X) g_{\mu \nu} + \B(\phi,X) \phi_{\mu} \phi_{\nu} \,,
\end{equation}
where $X = -g^{\mu\nu} \phi_\mu \phi_\nu / 2$, 
with $\phi_{\mu} = \p_\mu \phi$, 
whereas $\A(\phi,X)$ and $\B(\phi,X)$ are arbitrary functions of the 
scalar field $\phi$ and $X$. 
One can verify that the inverse metrics are related via
\begin{equation} \label{eq:inv_metric}
	\gtilde^{\mu\nu} = 
	\frac{1}{\A(\phi,X)}\left(g^{\mu\nu}-\frac{\B(\phi,X)}{\A(\phi,X)-2\B(\phi,X)X}{\phi^{\mu}\phi^{\nu}}\right) \,,
\end{equation}
where $\phi^{\mu} = g^{\mu\nu}\phi_{\nu}$,
such that 
$\gtilde^{\mu\nu} \gtilde_{\nu\alpha} = \delta^{\mu}_{~\alpha}$.

In  Jordan frame, the metric is minimally coupled to the matter 
fields, here collectively represented by $\Psi$, and the matter 
action is given by 
$S_{M}[\gtilde_{\mu \nu}, \Psi]$.
On the other hand, in the Einstein frame, which is obtained via the 
transformation eq.~\eqref{eq:metric_transf}, the metric is non-
minimally coupled to the matter fields. \footnote{We assume that the disformal couplings  $\A$ and $\B$ are the same for all matter species. We comment on the case of 
the non-universal disformal coupling in section \ref{subsec:non-universal}.}
More concretely, the action in the Einstein frame is given by
\begin{equation} \label{eq:einstein_frame}
	S[g_{\mu \nu}, \phi, \Psi] 
	    = \int d^4x \, \sqrt{-g} \left[ \frac{R}{16\pi G_*}  + K(\phi, X) \right] 
	    + S_{M}[\A g_{\mu \nu} + \B \phi_{\mu} \phi_{\nu}, \Psi] \,,
\end{equation}
where $G_*$ is the bare gravitational constant and $K(\phi, X)$ is an 
arbitrary function of $\phi$ and $X$. 
The determinant of $g_{\mu \nu}$ is related to the determinant of 
$\gtilde_{\mu \nu}$ as~\cite{Zumalacarregui:2012us}
\beqa
\sqrt{-\gtilde}=\A^{3/2}(A-2\B X)^{1/2}\sqrt{-g} \,, 
\label{determinant}
\eeqa
which can be derived by moving to a local Lorentz frame, in which 
$\phi=\phi(t)$~\cite{bekenstein}. 

In order to go back to Jordan frame, we employ the inverse metric 
transformation 
\begin{equation} 
	g_{\mu \nu} = \tilde{\A}(\phi,\Xtilde) \gtilde_{\mu \nu} + \tilde{\B}(\phi,\Xtilde) \phi_\mu \phi_\nu \,,
\end{equation}
where $\Xtilde = -\gtilde^{\mu\nu}\phi_\mu\phi_\nu / 2$,  
$\tilde{\A} = 1/{\A}$ and  $\tilde{\B} = -{\B}/{\A}$.  
For the resulting Einstein-Hilbert action in  Jordan frame, see 
refs.~\cite{Zumalacarregui:2013pma, Domenech:2015tca}.  

\subsection{Particle motion}

The trajectory of a test particle is parametrized by its proper time. 
The proper time in  Jordan frame, $\tiltau$, is related to the proper 
time in the Einstein frame, $\tau$, as
\begin{equation}
d{\tilde{\tau}}^2 =-\tilde{g}_{\mu\nu} d x^{\mu} d x^{\nu}
=-(\A g_{\mu\nu}+\B\phi_{\mu}\phi_{\nu}) d x^{\mu} d x^{\nu}
=\left(\A-\B(\phi_{\mu}u^{\mu})^2\right) d\tau^2\equiv {\alpha^2 d \tau^2} \,,
\label{propertime}
\end{equation}
where we have introduced the Einstein frame four-velocity 
$u^{\mu} = d x^\mu / d \tau$ and the function 
$\alpha = \sqrt{ \A - \B(\phi_{\mu} u^{\mu})^2 }$. 

For a test particle with four-velocity 
$\tilde{u}^{\mu} \equiv d x^{\mu} / d \tilde{\tau}$ in  Jordan frame, 
the geodesic equation is given by
\begin{equation}\label{geod_jordan}
\tilde{u}^{\nu}\tilde{\nabla}_{\nu}\tilde{u}^{\mu}=0 \,.
\end{equation}  

Now, to compute the equations of motion in the Einstein frame, first note that the four-velocity in Jordan frame is related to the one in Einstein frame as
\begin{equation} \label{eq:four_velocity}
	\tilde{u}^{\mu}  
		= \frac{d x^{\mu}}{d \tilde{\tau}}
		=\alpha^{-1}u^{\mu} \,.
\end{equation}
Also, the relation between the covariant derivatives 
$\tilde{\nabla}_{\mu}$ and $\nabla_{\mu}$ is given by~\cite{wald}
\beqa \label{eq:cov_dif_rel}
\tilde{\nabla}_{\nu}u^{\mu}=\nabla_{\nu}u^{\mu}+{C^{\mu}}_{\nu\rho}u^{\rho} \,,
\eeqa
such that, for the metric transformation eq.~\eqref{eq:metric_transf}, one can show that~\cite{Domenech:2015tca}
\begin{align} \label{eq:connection}
{C^{\mu}}_{\nu\rho}
	&= \frac12 \gtilde^{\mu\sigma}\left(\nabla_{\rho}\gtilde_{\sigma\nu}+
		\nabla_{\nu}\gtilde_{\sigma\rho}-\nabla_{\sigma}\gtilde_{\nu\rho}\right) \\
	&=  
		\delta^\mu_{~(\nu} \nabla_{\rho )} \ln \A 
		- \frac{1}{2} \gtilde_{\nu\rho} \gtilde^{\mu \sigma} \nabla_\sigma \ln \A
		+ \frac{\B}{\A-2X\B} \phi^\mu \phi_{\nu\rho} \nonumber \\
		&\hspace{0.4cm}+ \frac{\B}{\A-2X\B} \phi^\mu \phi_{(\nu} \nabla_{\rho )} \ln (\B/\A)
		- \frac{\B}{2} \phi_\nu \phi_\rho \, \gtilde^{\mu \sigma} \nabla_\sigma \ln (\B/\A) \,, 		
\end{align}
where $\phi_{\mu\nu} = \nabla_\mu \nabla_\nu \phi$. 
Note also that the metric $\gtilde_{\mu\nu}$ and its inverse are to 
be computed using eqs.~\eqref{eq:metric_transf} and~\eqref{eq:inv_metric}, respectively. 
Combining these results, the geodesic equation~\eqref{geod_jordan} can be written in terms of Einstein frame quantities, yielding
\begin{eqnarray}\label{geod_ein}
	u^{\nu}\nabla_{\nu}u^{\mu} 
		= {\alpha^{-1}u^{\mu}u^{\nu}\nabla_{\nu}\alpha}
		-{C^{\mu}}_{\nu\rho}u^{\nu}u^{\rho} \,,
\end{eqnarray} 
which is the equation of motion for a test particle in Einstein 
frame.
Since the matter fields are non-minimally coupled to gravity in the $g_{\mu\nu}$ frame, the test particle does not follow geodesics.

\subsection{Geometric optics and propagation of photons}

To derive the geodesic equation for photons, we consider the Maxwell 
equations in the geometric optics approximation~\cite{mtw}.  
From the action for electromagnetism in Jordan frame 
\beqa 
S_{EM} = -\frac14 \int d^4 x\sqrt{-\gtilde} \,
	\gtilde^{\alpha\mu} \gtilde^{\beta\nu} F_{\mu\nu} F_{\alpha\beta} \,,
\label{action:photon}
\eeqa
where $F_{\mu\nu}=\partial_{\mu}A_{\nu}-\partial_{\nu}A_{\mu}$ 
with $A_{\mu}$ being the electromagnetic gauge field, 
the source-free Maxwell equations, in the Lorentz gauge, are given by
\beqa
\label{maxwell}
\nablatilde^{\nu}\nablatilde_{\nu}A_{\mu}-{\Rtilde_{\mu}}^{~~\nu}A_{\nu}&=&0 \,,\\
\label{gauge}
\nablatilde^{\mu}A_{\mu}&=&0 \,.
\eeqa
Under the geometric optics approximation, we seek a solution of the 
form
\beqa
A_{\mu}=C_{\mu}e^{i S} \,, 
\label{geometricoptics}
\eeqa
where we assume that the phase $S$ is large, varying rapidly compared 
to the amplitude $C_{\mu}$, and that the wavelength of the 
electromagnetic waves is much smaller than the curvature radius. 
Substituting eq.~(\ref{geometricoptics}) into eq.~(\ref{maxwell}) 
yields, to the leading order 
\beqa
\gtilde^{\mu\nu}k_{\mu}k_{\nu}=0 \,,\label{nullcondition}
\eeqa
where $k_{\mu}=\partial_{\mu}S$ is the propagation vector of the 
electromagnetic waves, which is normal to the surfaces of constant 
phase. 
This shows that, in Jordan frame, the wave vector 
$\tilk^\mu = \gtilde^{\mu\nu}k_{\nu}$ is a null vector.
The geodesics equation is derived by taking the gradient of 
eq.~(\ref{nullcondition})
\beqa
\tilk^{\nu}\nablatilde_{\nu}k_{\mu}=0 \,.
\eeqa
Thus, the light rays travel along null geodesics.

In the Einstein frame, the propagation vector 
$k_{\mu}=\partial_{\mu}S$ is the same, but its norm is different from 
zero: from eqs.~(\ref{nullcondition}) and~(\ref{eq:inv_metric}),
\beqa
g^{\mu\nu}k_{\mu}k_{\nu}=\frac{\B}{\A-2\B X}(\phi^{\mu}k_{\mu})^2 \,.
\label{null:e-frame}
\eeqa
Then, by taking the gradient of eq.~(\ref{null:e-frame}), the 
geodesics equation is given by
\beqa
k^{\nu}\nabla_{\nu}k_{\mu}=\frac12\nabla_{\mu}
\left(\frac{\B(\phi^{\mu}k_{\mu})^2}{\A-2\B X}\right) \,,
\label{geodesics:e-frame}
\eeqa
where $k^{\mu}=g^{\mu\nu}k_{\nu}$. 

The fact that the tangent vector $k_{\mu}$ is no longer null in the 
Einstein frame (although being null in Jordan frame) implies that the 
speed of light in the $g_{\mu\nu}$ frame is different from unity. 
In fact, for a flat spacetime in the Einstein frame, light rays obey
\beqa
0=d\tils^2=\A ds^2+\B\dot\phi^2dt^2=-\alpha^2dt^2+\A d{\bf x}^2 \,,
\eeqa
where we assume $\phi(t,{\bf x})=\phi(t)$ and 
$\alpha=\sqrt{\A-\B\dot\phi^2}$. 
Thus, the effective speed of light in the Einstein frame $c_E$ is 
given by
\beqa
c_E=\frac{\alpha}{\A^{1/2}} \,.
\label{speedoflight}
\eeqa

\subsection{Newtonian limit and gravitational constant}

We now turn our attention to the Newtonian limit (weak gravitational 
field and slow motion of test particles).
In standard general relativity, the geodesic equation (for minimally 
coupled particles) can yield the Newtonian limit for a weak 
gravitational field only if 
\begin{equation}
    g_{00} = -(1 + 2\Phi) \,, \quad
    g_{ij} = \delta_{ij} \,,
\end{equation}
where $\Phi$ is the Newtonian potential.
In accordance with the post-Newtonian bookkeeping system~\cite{will2018theory},
we assign a bookkeeping label $\epsilon$ that keeps track of small 
quantities such that 
$|\partial/\partial t|/|\partial/\partial x|\sim v$
and $ \Phi \sim v^2 \sim p/\rho \sim \partial \phi \sim\,\,       \order{\epsilon}$, 
where $v$ represents the velocity of test particles.
We naturally follow the same approach for quantities in  Jordan frame.

Note however that since the disformal factor $\B$ is always 
accompanied by factors 
$(\p\phi)^2 \sim \order{\epsilon^2}$, 
these terms will lead to post-Newtonian corrections.
Hence, as it will be shown, the Newtonian limit of disformally 
related theories is identical to the conformal case with $\A = \A(\phi)$.
\footnote{For the analysis of the post-Newtonian order of 
the metric, see~\cite{Ip:2015qsa}.}

Let us start with the Newtonian limit in Einstein frame.
For a canonical scalar field, $K(\phi, X) = X - V(\phi)$, the action 
eq.~\eqref{eq:einstein_frame} leads to the equation of motion
\begin{equation}
    \Box \phi - \frac{dV}{d\phi} = Q \,,
\end{equation}
where $Q$ represents the non-minimal coupling with matter fields and 
is given by~\cite{Chibana:2019jrf}
\begin{align}
    Q &= \nabla_\mu W^\mu - Z \,, 
\\
    W^\mu &= \frac{\B}{\A} T^{\mu\nu}_{M} \phi_\nu 
        - \frac{\A - 2\B X}{2\A (\A - \A_{,X} + 2 \B_{,X} X^2)}
            \left( \A_{,X} T_{M} + \B_{,X} T^{\alpha\beta}_{M} \phi_\alpha \phi_\beta \right) \phi^\mu \,, 
\\
    Z &= \frac{1}{2\A} 
        \Biggl( 
            \qty[ \A_{,\phi} + \frac{\A_{,X} X \left(\A_{,\phi} - 2 \B_{,\phi} X \right)}{\A - \A_{,X} + 2 \B_{,X} X^2} ] T_{M} 
\\
    &\phantom{= \frac{1}{2\A} \Biggl(}
                + \qty[ \B_{,\phi} + \frac{\B_{,X} X \left(\A_{,\phi} - 2 \B_{,\phi} X \right)}{\A - \A_{,X} + 2 \B_{,X} X^2} ]
                T^{\mu\nu}_{M} \phi_\mu \phi_\nu
            \Biggr) \,,
\end{align}
$T^{\mu\nu}_{M}$ is the matter energy-momentum tensor and  
$T_{M}$ is its trace and the comma denotes the functional derivative such that 
$\A_{,X}=\partial \A/\partial X$, for instance.
Hereafter we set $V(\phi)=0$ for simplicity.
Then, keeping only terms of order $\order{\epsilon}$, the right-hand 
side of the equation of motion becomes
\begin{equation} 
    Q = \frac{1}{2} \pdv{\ln \A}{\phi}\rho  \,.
\end{equation}
Hence, in the Newtonian limit one has
\begin{equation} \label{eq:eom_phi_ef}
    \laplacian{\phi} = \frac{1}{2} \pdv{\ln\A}{\phi} \rho \,,
\end{equation}
which admits the solution, up to order $\order{\epsilon}$, 
\begin{equation} \label{eq:phi_ef}
    \phi(\vb{x}) 
        = \phi_0 - \frac{1}{8\pi} \eval{\pdv{\ln\A}{\phi}}_{\phi_0}
            \int d^3 x'\frac{\rho(\vb{x'})}{\abs{\vb{x} - \vb{x'}}} \,,
\end{equation}
where the derivative of $\A$ is evaluated at $\phi_0$, the asymptotic 
value of the scalar field. 

On the other hand, the evolution of the metric $g_{\mu\nu}$ is 
determined from eq.~\eqref{eq:einstein_frame}, leading to the usual 
Einstein's equation,
\begin{equation} \label{eq:eom_g_ef}
R_{\mu\nu} = 8\pi G_* \left[ T_{\mu\nu}^{M} + T_{\mu\nu}^{\phi}
-\frac12 g_{\mu\nu} \left(T^{M} + T^{\phi} \right) \right] \,,
\end{equation}
in which $T^{\phi}$ is the trace of $T_{\mu\nu}^{\phi}$, the energy-momentum tensor for the canonical scalar field, given by
\begin{equation}
    T_{\mu\nu}^{\phi} = \phi_\mu \phi_\nu + g_{\mu\nu} X \,.
\end{equation}
We note that $T_{\mu\nu}^\phi$ is $\order{\epsilon^2}$, and in the 
Newtonian limit, the time-time component of eq.~\eqref{eq:eom_g_ef}
becomes 
\begin{equation}
    \laplacian{\Phi} = 4\pi G_*\rho \,.
\end{equation}
Similar to $\phi$, the solution for the gravitation potential is
\begin{equation} \label{eq:Phi_ef}
    \Phi(\vb{x}) 
        = \Phi_0 - G_*
            \int d^3 x'\frac{\rho(\vb{x'})}{\abs{\vb{x} - \vb{x'}}} \,.   
\end{equation}

To find the gravitational constant in Einstein frame, we use the fact 
that due to the non-minimal coupling to the scalar field, material 
particles experience a fifth-force. 
At the Newtonian order, the geodesic equation~\eqref{geod_ein} 
becomes
\begin{align}
    \frac{d^2 \vb{x}}{dt^2} 
        &= - \grad{\Phi} - \frac{1}{2} \pdv{\ln\A}{\phi} \grad{\phi}  \\
        &= -G_*
            \qty[ 1 + \frac{1}{16\pi G_*} \eval{\qty(\pdv{\ln\A}{\phi})^2}_{\phi_0} ]
            \int d^3 x' \frac{ \qty(\vb{x} - \vb{x'}) \rho(\vb{x'})}{\abs{\vb{x} - \vb{x'}}^3} \,,
\end{align}
where we have employed eqs.~\eqref{eq:phi_ef} and~\eqref{eq:Phi_ef}.
The coefficient before the integral can be seen as the effective gravitational constant in Einstein frame measured by Cavendish-type experiments, and thus 
\begin{equation}
    G = G_{*} \qty[ 1 + \frac{1}{16\pi G_*} \eval{\qty(\pdv{\ln\A}{\phi})^2}_{\phi_0} ]
    \,.
\end{equation}

Now we move on to Jordan frame.
To calculate the effective gravitational constant $\tilG$ we must 
first get the corresponding equations of motion, 
eqs.~\eqref{eq:eom_phi_ef} and~\eqref{eq:eom_g_ef}, in  Jordan frame.

Ignoring post-Newtonian terms, the connection relating the covariant 
derivatives in each frame, eq.~\eqref{eq:connection}, takes the form 
\begin{equation}
    {C^\rho}_{\mu\nu} 
        = \frac{1}{2} \pdv{\ln\A}{\phi}
        \qty( \delta^\rho_\mu \phi_\nu + \delta^\rho_\nu \phi_\mu 
            - \tilg_{\mu\nu} \tilg^{\rho\sigma} \phi_\sigma ) \,,
\end{equation}
being a $\order{\epsilon}$ quantity, 
and employing eq.~\eqref{eq:cov_dif_rel} one finds 
$\Box\phi = \A \tilbox \phi$.
Thus, the equation of motion for $\phi$, eq.~\eqref{eq:eom_phi_ef}, 
written in terms of Jordan frame quantities becomes
\begin{equation} \label{eq:eom_phi_jf}
    \tilnab^2\phi = \frac{1}{2} \pdv{A}{\phi} \tilrho \,,
\end{equation}
where we have used eq.~\eqref{T_correspondence} to relate $\rho$ to  
$\tilrho$.

Now we calculate the equation of motion for $\tilg_{\mu\nu}$ from its 
Einstein-frame counterpart.
For the metric transformation eq.~\eqref{eq:metric_transf} one can 
show the relation~\cite{Domenech:2015tca}
\begin{equation}
    R_{\mu\nu} 
        = \tilR_{\mu\nu} - \tilnab_\rho {C^\rho}_{\mu\nu}
        - {C^\rho}_{\mu\sigma} {C^\sigma}_{\rho\nu}
        + \tilnab_\mu \tilnab_\nu \ln \sqrt{\frac{- \tilg}{-g}}
        + {C^\rho}_{\mu\nu} \tilnab_\rho \ln \sqrt{\frac{- \tilg}{-g}} \,,
\end{equation}
such that, neglecting terms $\sim \order{\epsilon^2}$, 
eq.~\eqref{eq:eom_g_ef} becomes
\begin{equation}
\tilR_{\mu\nu} = 8\pi G_* \A \left(\tilT_{\mu\nu}-\frac12 g_{\mu\nu} \tilT \right) 
        - \pdv{\ln\A}{\phi} 
        \qty( \tilnab_\mu \tilnab_\mu \phi + \frac12 \tilg_{\mu\nu} \tilbox\phi ) \,,
\end{equation}
where $\tilT$ is the trace of $\tilT_{\mu\nu}$. 
For the metric in the Newtonian limit, $\tilg_{00} = -(1+\tilPhi)$, 
$\tilg_{ij} = \delta_{ij}$, the time-time component of the above 
equation becomes
\begin{equation}
    \laplacian{\tilPhi} 
        = 4\pi G_* \eval{\A}_{\phi_0}  
        \qty[ 1 + \frac{1}{16\pi G_*} \eval{\qty(\pdv{\ln\A}{\phi})^2}_{\phi_0} ]
        \tilrho
\end{equation}
where we have used eq.~\eqref{eq:eom_phi_jf}.
The Newtonian potential is related to $\tilrho$ via 
$\laplacian{\tilPhi} = 4\pi \tilG \tilrho$, so we identify the effective gravitational constant in Jordan frame:
\begin{equation}
     \tilG = \eval{\A}_{\phi_0}  G_{*} \qty[ 1 + \frac{1}{16\pi G_*}
     \eval{\qty(\pdv{\ln\A}{\phi})^2}_{\phi_0} ] \,.
\end{equation}

Hence, the effective gravitational constants are related as 
$\tilG = \eval{\A}_{\phi_0}  G $, 
which is the same as in the conformal case~\cite{cy}.
As mentioned above, this result was expected since the disformal term 
leads to post-Newtonian corrections.

\section{Cosmology}
\label{sec:cosmology}

In this section, we study several cosmological observables and, for 
each of them, provide the correspondence between Jordan-frame and 
Einstein-frame quantities.

\subsection{Redshift}

As the first example of cosmological interest, let us examine the 
redshift, and check whether it is invariant under disformal metric 
transformations or not, although the invariance has already been 
shown by~\cite{Domenech:2015hka}. 
The line elements of the Friedmann-Robertson-Walker (FRW) spacetime 
in the Jordan frame and Einstein frame are given respectively by
\beqa
&&d \tils^2 = -d \tilt^2 + \tila(\tilt)^2 \left(d\chi^2+\sin^2_K\chi d\Omega^2\right) \,, \\
&&ds^2=-dt^2+a(t)^2 \left(d\chi^2+\sin^2_K\chi d\Omega^2\right) \,,
\eeqa
where $\tilt$ is the cosmic time, $\tila$ is the scale factor, and 
$\sin_K\chi=\sin\chi, \chi,\sinh\chi $ for an open, flat, or closed 
universe, respectively.
Since  $d \tils^2=\A d s^2+\B\dot\phi^2 d t^2$, the corresponding 
cosmic time and scale factor in Einstein frame are given by
\beqa
d \tilt=\sqrt{\A-\B\dot\phi^2} d{t}=\alpha d{t} \,,
\quad
\tila(\tilt)={\A}^{1/2}a(t) \,.
\label{tau:scale}
\eeqa

First, we calculate the observed redshift in Jordan frame.  
Consider a light pulse emitted with a frequency $\tilomega_S$ by a 
source, and measured with a frequency $\tilomega_O$ by an observer.
Here and in what follows, the subscripts $S$ and $O$ refer to 
quantities measured at the position of the source and observer, 
respectively.
Then, since photons are minimally coupled to the metric in Jordan 
frame,  the redshift is given by
\beqa \label{eq:redshift_jordan}
1+\tilz=\frac{\tilomega_S}{\tilomega_O}=\frac{\tila_O}{\tila_S} \,. 
\eeqa 

Next, we turn to the redshift in Einstein frame. 
To do so, one must be careful about which units are being used to 
measure the frequency~\cite{cy}.
That is because the observation is made in the observer's reference 
frame, in which the units are, in general, different from those in 
the source's reference frame.
In particular, according to eq.~\eqref{tau:scale}, for the same 
frequency $\tilomega$ in  Jordan frame, the corresponding frequency 
in Einstein frame, as measured at a given position, would be 
$\omega_1 = \alpha_1 \tilomega$, whereas when measured at another, different position, would result in $\omega_2 = \alpha_2 \tilomega$.
The relation between these measurements is  
\beqa
\frac{\omega_1}{\omega_2}=\frac{\alpha_1}{\alpha_2} \,,
\label{relation}
\eeqa
corresponding to the change in the unit of time from one spacetime 
point to another.
\footnote{Using the effective speed of light 
eq.~(\ref{speedoflight}), given by $c_E=\alpha/\A^{1/2}$, the 
corresponding relation between two wavelengths, $\lambda_1$ and 
$\lambda_2$, is $\lambda_1/\lambda_2=\A_2^{1/2}/\A_1^{1/2}$, which 
translates to a change in the unit of length.}

Therefore, the observed redshift in Einstein frame is, more 
precisely, defined as the ratio between the frequency at the source, 
measured in units of the observer's reference frame, and the observed 
frequency. 
Taking into account the relation eq.~(\ref{relation}), it reads
\beqa
1+z=\frac{\alpha_O}{\alpha_S}\frac{\omega_S}{\omega_O}=\frac{\tilomega_S}{\tilomega_O}=1+\tilz \,.
\label{redshift:einstein}
\eeqa 
Thus, the observed redshift is disformal-frame independent~\cite{Domenech:2015hka}. 
Furthermore, in agreement with eqs.~\eqref{tau:scale} 
and~\eqref{redshift:einstein}, the relation between the redshift and 
scale factor in Einstein frame is given by
\begin{equation} \label{eq:z_scale_einstein}
	1+z = \sqrt{\frac{\A_O}{\A_S}} \frac{a_O}{a_S} \,.
\end{equation}

\subsection{Redshift drift}

The redshift drift (also known as Sandage-Loeb effect~\cite{Sandage:1962,Loeb:1998bu})
is the change of the redshift of the source after some time interval.  
It has been said that the redshift drift can be a direct probe of the 
cosmic acceleration ~\cite{Corasaniti:2007bg}.   
To see this, let us consider light emitted at $\tilt_S$ and at 
$\tilt_S+\Delta \tilt_S$ from a source and observed at $\tilt_O$ and 
at $\tilt_O+\Delta \tilt_O$ by an observer. 
The redshift change is given by
\beqa
\Delta \tilz=\frac{\tila(\tilt_O+\Delta \tilt_O)}{\tila(\tilt_S+\Delta \tilt_S)}-\frac{\tila(\tilt_O)}{\tila(\tilt_S)}
\simeq   \frac{\dot \tila(\tilt_O)-\dot \tila(\tilt_S)}{\tila(\tilt_S)}\Delta \tilt_O \,,
\eeqa
where a dot denotes the derivative with respect to $\tilt$ and we  
used the relation 
$\Delta \tilt_O/\tila(\tilt_O)=\Delta \tilt_S/\tila(\tilt_S)$ 
and assumed the time interval is small in the second equality.  
Therefore, $\Delta \tilz>0 (<0)$ corresponds to 
$\dot \tila(\tilt_O)-\dot \tila(\tilt_S) >0 (<0)$, 
which implies the accelerating (decelerating) universe. 
The redshift drift is a kinematical relation and makes no use of the 
Friedmann equation. 
As such, the detection of a positive sign of $\Delta \tilz$ could be 
a direct proof of the acceleration of the cosmic expansion. 

However, this statement is frame-dependent because, according to 
eq.~\eqref{tau:scale}, the scale factor in Einstein frame involves 
$\A$.
More concretely, $\Delta z$ in Einstein frame is given by
\begin{equation}
\Delta z = \Delta \tilz
        = \frac{1}{ \eval{\A^{1/2} \, a(t)}_S} 
            \qty( \frac{1}{\alpha}\frac{d}{d t} \eval{\qty[ \A^{1/2} \, a(t) ] }_O
                - \frac{1}{\alpha}\frac{d}{d t} \eval{\qty[ \A^{1/2} \, a(t) ] }_S ) \eval{ \alpha \Delta t }_O \,.
\end{equation} 
Therefore, the sign of $\Delta z$ in Einstein frame has no direct 
connection to the cosmic acceleration.

\subsection{Luminosity distance and Hubble's law}

The luminosity distance in Jordan frame $\tild_L$ is defined in terms 
of the source's (absolute) luminosity, $\tilL_S$, and the observed 
flux, $\tilf_O$, as $\tild_L=\sqrt{\tilL_S/4\pi \tilf_O}$. 
Due to the expansion of the universe, $\tilL_S$ is redshifted as 
$\tilL_O=\tilL_S(\tilomega_O/\tilomega_S)^2$, 
where $\tilL_O$ is the apparent luminosity measured by the observer.
On the other hand, the observed flux is defined as 
$\tilf_O=\tilL_O/4\pi \tila_O^2\sin^2_K\chi$, 
where $\chi$ is the comoving distance to the source,
\beqa
	\chi = \int^{\tilt_O}_{\tilt_S}\frac{d{\tilt}}{\tila}
		   =\int^{\tilz}_{0} \frac{d{\tilz}}{\tila_O\tilH(\tilz)} \,.
\label{chi}
\eeqa
As a result, the luminosity distance in  Jordan frame takes the usual 
form
\beqa
\tild_L=\tila_O(1+\tilz)\sin_K\chi \,.
\label{dL}
\eeqa

Also, by taking the low-$\tilz$ limit,  we obtain Hubble's 
law,
\beqa
\tild_L=\frac{\tilz}{\tilH_0} \,,
\eeqa
where $\tilH_0=\tilH(0)$ is the present Hubble parameter. 

Now we calculate the luminosity distance $d_L$ in Einstein frame. 
According to the discussion regarding the redshift, to define the 
luminosity distance one should consistently use the units in the 
observer's reference frame.
Since the luminosity has units of energy divided by time, the 
source's absolute luminosity, when measured in the observer's units, 
becomes $L_S (\alpha_O/\alpha_S)^2$.
Also, $L_S$ is related to the apparent luminosity, at the observer's 
position, as $L_S=L_O(\omega_S/\omega_O)^2$.
Therefore, the luminosity distance in Einstein frame is given by
\beqa \label{eq:d_L_einstein}
	d_L 
		= \sqrt{\frac{L_S}{4\pi f_O}\left(\frac{\alpha_O}{\alpha_S}\right)^2}
		= a_O(1+z)\sin_K\chi \,.
\eeqa
Thus, as in the case of the conformal transformation~\cite{cy}, 
$d_L$ differs from $\tild_L$ by the conformal factor:
$d_L=\tild_L/\A^{1/2}_O$.  
Note, however, that, due to the disformal transformation, $\chi$ is 
no longer determined by the null geodesics in the Einstein frame. 

Once again, by taking the low-$z$ limit, we obtain the 
Hubble's law in Einstein frame,
\beqa
d_L=\frac{z}{\A^{1/2}_O\tilH_0}\equiv \frac{z}{H_{E0}},
\label{hubble-e}
\eeqa
where $H_E=\A^{1/2}\tilH$ is the effective Hubble parameter,\footnote{Here we use $z$ to define the effective Hubble parameter as eq.~(\ref{hubble-e}).  If we use the recession velocity $v$, we need to multiply $z$ by the effective speed of light $c_E=\alpha/\A^{1/2}$  
so that $v=c_E z=\alpha \tilH d_L$, and the effective Hubble parameter is then 
$\alpha\tilH$. }
\beqa
H_E&=&\A^{1/2}\frac{1}{\A^{1/2}a}\frac{1}{\alpha}\frac{d}{dt}\left(\A^{1/2}a\right)\nonumber\\
&=&\frac{\A^{1/2}}{\alpha}\left(H+\frac12\frac{d\A/dt}{\A}\right),
\eeqa
and $H = (1/a)(d a / d t)$ is the usual one.

\subsection{Horizon problem and inflation}

Let us discuss the condition for solving the horizon problem in both 
frames. 
Since the Universe became almost homogeneous right after the 
decoupling epoch, the horizon problem can be solved if the distance 
light travels between two times (say, $\tilt_i$ and $\tilt_f$, with 
$\tilt_f < \tilt_{\rm dec}$) is larger than the current size of the 
horizon. In comoving coordinates, this condition in Jordan frame can 
be expressed as
\beqa
	\int_{\tilt_i}^{\tilt_f} \frac{d{\tilt}}{\tila} 
		 =\int_{\tila_i}^{\tila_f} d{\ln\tila'} \frac{1}{\tila' \widetilde{H}(\tila')}
		 > \frac{1}{\tila_0 \widetilde{H}_0} \,.
\label{eq:HcondJ}
\eeqa
From $\tila_0 \widetilde{H}_0 = a_0 H_{E0}$ and 
eq.~\eqref{tau:scale}, this condition can be recast into
\beqa
 \int_{t_i}^{t_f} \frac{\alpha\,d{t}}{\A^{1/2}\,a} 
 	> \frac{1}{a_0 H_{E0}} \,.
\label{eq:HcondE}
\eeqa

Since the comoving Hubble distance, $(\tila \widetilde{H})^{-1}$,
increases with time in the standard Big-Bang cosmology, it is 
manifest from eq.~(\ref{eq:HcondJ}) that, in order to solve the 
horizon problem, there must be some period in the past, during which 
$(\tila \widetilde{H})^{-1}$ decreased, that is,
\beqa 
	\frac{d}{d\tilt} \left( \frac{1}{\tila \widetilde{H}} \right)
  		= - \left( \frac{d\tila}{d\tilt} \right)^2 \frac{d^2\tila}{d\tilt^2} < 0 \,.
\eeqa
This condition is exactly the same as that of inflation in the
expanding Universe. 

In Einstein frame, thanks to the relation 
$\tila \widetilde{H} = a H_{E}$, this condition can be rewritten as
\beqa
 \frac{d}{dt} \left( \frac{1}{a H_E} \right) 
= \frac{d}{dt} \left( \frac{\alpha}{\A^{1/2}} 
    \frac{1}{\frac{da}{dt}+a \frac{d}{dt}(\ln \A^{1/2})} \right) 
< 0 \,,
\eeqa
assuming $\alpha > 0$.  
Therefore, as in the case of the redshift drift, the cosmic 
acceleration does not have a direct connection to the solution of the 
horizon problem.~\footnote{See~\cite{Clayton:1998hv, Magueijo:2003gj, Kaloper:2003yf} for earlier attempts to solve cosmological puzzles within the disformal theories.}

\subsection{Angular diameter distance and the distance-duality relation}

Now we turn our attention to the angular diameter distance.
Consider an object with proper diameter $\tilD$, subtending an angle 
$\theta$, as measured by an observer.
In  Jordan frame, the angular diameter distance to said object, 
$\tild_A$, is defined as
\beqa
 \widetilde{d}_A \equiv \frac{\widetilde{D}}{\theta} 
  = \widetilde{a}_S \sin_K\chi
  = \frac{1}{{1+}\tilz} \widetilde{a}_O \sin_K\chi \,.
\eeqa
As a result of the conservation of the number of photons~\cite{ellis}, 
the angular diameter distance is related to luminosity distance, 
eq.~\eqref{dL}, through the so-called distance-duality relation, 
$\tild_L = (1+\tilz)^2 \tild_A$.

On the other hand, the proper diameter $D$ in Einstein frame, in 
units of the observer's position, becomes $D\sqrt{\A_S/\A_O}$, where 
the factor $\sqrt{\A_S/\A_O}$ comes from the change in the unit of 
length from the source's reference frame to the observer's reference 
frame.  
Hence, the angular diameter distance in Einstein frame is given by
\beqa \label{eq:d_A_einstein}
  d_A 
  	= \sqrt{\frac{\A_S}{\A_O}} \frac{D}{\theta} 
  	= \sqrt{\frac{\A_S}{\A_O}} a_S \sin_K\chi
  	=\frac{a_O}{1+z}\sin_K\chi \,,
\eeqa 
where we have used eq.~\eqref{eq:z_scale_einstein} to replace the 
scale factor $a_S$ in favor of the redshift.
Once again, as in the case of the luminosity distance, $d_A$ differs 
from $\tild_A$ by the conformal factor, $d_A=\tild_A/\A^{1/2}_O$, 
such that the distance-duality relation $d_L = (1+z)^2 d_A$ still 
holds.\footnote{
This result seems to be different from the one in 
ref.~\cite{brax}. 
The difference comes from the definition of the redshift in Einstein 
frame, which in~\cite{brax} is done without the unit conversion 
factor given in eq.~(\ref{redshift:einstein}), and hence, is not 
frame-independent quantity. 
When the unit conversion is taken into account, the result from~\cite{brax} agrees with ours. }
This stems from the fact that the electromagnetic gauge 1-form is invariant under disformal transformation~\cite{Domenech:2015hka}, and hence, the (modified) 
flux conservation law for photons still holds true in the Einstein frame.  
In the following subsection, we investigate the distance-duality relation in general spacetimes and show its frame-invariance.

\subsection{Reciprocity relation and distance-duality relation in general spacetimes}
\label{sec:rec_relation}

We can derive  the distance-duality relation, the relation between the luminosity distance and the angular diameter distance,  in general spacetime 
in the Einstein frame following~\cite{ellis,sasaki}. We first derive the (modified) 
reciprocity relation, the relation between two geometrical distances~\cite{ellis}.  
Before doing that, in order to introduce the notation, we first explain  
the reciprocity relation in  Jordan frame.

\subsubsection{Reciprocity relation in  Jordan frame}

In  Jordan frame the second leading order of eq.~(\ref{maxwell}) gives
\beqa
\nablatilde_{\mu}(C^2\tilk^{\mu})=0 \,,
\label{amplitude}
\eeqa
where $C^2=\gtilde^{\mu\nu}C_{\mu}C_{\nu}$. Consider a null geodesics congruence diverging from a source with cross-sectional area $d\tilS$ perpendicular 
to the propagation vector $k_{\mu}$. Then the rate of change of $d\tilS$ along the null geodesics is determined by~\cite{mtw}
\beqa
\tilk^{\mu}\nablatilde_{\mu}d\tilS=d\tilS\nablatilde_{\mu}\tilk^{\mu} \,.
\label{expansion}
\eeqa
Therefore, from  eq.~(\ref{amplitude}) and eq.~(\ref{expansion}), 
\beqa
C^2d\tilS={\rm constant}
\label{conserved}
\eeqa
along the geodesics.  We note that the gauge condition eq.~(\ref{gauge}) implies 
$\gtilde^{\mu\nu}C_{\mu}k_{\nu}=0$ which ensures that $C_{\mu}$ is spacelike. 

The energy-momentum tensor of light rays constructed out of the action 
eq.~(\ref{action:photon}) under the geometric optics approximation is given by
\beqa
\Ttilde^{\mu\nu}=C^2\tilk^{\mu}\tilk^{\nu} \,. 
\eeqa
Then the energy flux measured in the rest frame of an observer with four-velocity 
$\tilu^{\mu}$ is~\cite{ellis,sasaki}
\beqa
\tilf^{\mu}=-\tilh^{\mu\nu}\Ttilde_{\nu\rho}\tilu^{\rho}=\tilf \tiln^{\mu} \,,
\label{fluxvector}
\eeqa
where the vector $\tiln^{\mu}$ is the unit vector in the spatial direction of the propagation vector seen 
on the rest frame of the observer projected by $\tilh^{\mu\nu}$ and $\tilf$ is the observed flux 
that are given by 
\begin{subequations}
\beqa
\tilh^{\mu\nu}&=&\gtilde^{\mu\nu}+\tilu^{\mu}\tilu^{\nu} \,, \\
\tilk^{\mu}&=&(-\tilu_{\nu}\tilk^{\nu})(\tilu^{\mu}+\tiln^{\mu})=\tilomega(\tilu^{\mu}+\tiln^{\mu}) \,, \\
\tilf&=&C^2\tilomega^2 \,.
\label{flux}
\eeqa
\end{subequations}
  
Since the energy density measured by the observer is 
$\Ttilde_{\mu\nu}\tilu^{\mu}\tilu^{\nu}=C^2\tilomega^2$, eq.~(\ref{amplitude}) and 
hence eq.~(\ref{conserved}) implies 
the conservation of the number of photons along the bundle of light rays.

The luminosity of the source $\tilL_S$ is determined through the flux $\tilf_S$ measured on a small sphere 
around the source in the source rest frame with the radius $\Delta \tilr$ as 
\beqa
 \tilL_S=4\pi\Delta\tilr^2 \tilf_S =4\pi \Delta\tilr^2C^2_S\tilomega_S^2 \,. 
\eeqa
The luminosity distance $\tild_L$ is defined by $\tild_L=\sqrt{\tilL_S/4\pi \tilf_O}$. 

Let us consider the situation where we observe the flux from the source 
along some bundle of geodesics which subtends a solid angle $d\Omega_S$ at 
the source and has cross-sectional area $d\tilS_S$ at the observer. 
We define the source area distance $\tilr_S$ by~\cite{ellis}
\beqa
d\tilS_S=\tilr_S^2 d\Omega_S \,.
\eeqa
Note that $\tilr_S$ is not an observable because $d\Omega_S$ at 
the source cannot be measured.  However, $\tilr_S$ is related to 
the luminosity distance $\tild_L$. From the photon number conservation eq.~(\ref{conserved}),  we have $C_S^2\Delta\tilr^2d\Omega_S=C_O^2d\tilS_S$. Then
\beqa
\tild_L^2=\frac{\tilL_S}{4\pi \tilf_O}=\frac{C_S^2}{C_O^2}\frac{\tilomega_S^2}{\tilomega_O^2}\Delta\tilr^2=(1+z)^2\frac{d\tilS_S}{d\Omega_S}=(1+z)^2\tilr_S^2 \,.
\label{dl-rs}
\eeqa

Next consider the situation where we measure 
the solid angle $d\Omega_O$ subtended by some objects whose cross-sectional area 
is $d\tilS _O$. Then, we define the observer area distance $\tilr_O$ by~\cite{ellis}
\beqa \label{eq:ro_jf}
d\tilS_O=\tilr_O^2 d\Omega_O \,.
\eeqa
{}From this definition, it is clear that $\tilr_O$ is nothing 
but the angular diameter distance $\tild_A$. 
The relation between $\tilr_S$ and $\tilr_O$ is known as the reciprocity 
relation, which is written as
\beqa
\tilr_S=(1+z)\tilr_O \,. 
\label{reciprocity}
\eeqa
The reciprocity relation is derived from the null geodesics deviation 
equation~\cite{ellis,sasaki}. The relation between  $\tild_L$ and $\tild_A$ is 
sometimes called as the distance-duality relation, which is written using eq.~(\ref{dl-rs}) as
\beqa
\tild_L=(1+z)^2\tild_A \,.
\eeqa
The relation is a consequence of the photon number conservation.

\subsubsection{Reciprocity relation in Einstein frame}

In the Einstein frame, we work in the uniform $\phi$ gauge in which $\phi=\phi(t)$. 
Hereafter, in this subsection,  we assume $g^{0i}=0$, which is equivalent to $g^{00}=1/g_{00}$, for simplicity.
Then $2X=-g^{\mu\nu}\phi_{\mu}\phi_{\nu}=(\phi_{\mu}u^{\mu})^2$ for a comoving observer, 
such that $\alpha=\sqrt{\A-\B(\phi_{\mu}u^{\mu})^2}=\sqrt{\A-2\B X}$. 
Moreover, we assume $\phi(t)$ varies slowly compared with the phase $S$ 
in the geometric optics approximation and only consider the terms up to the first order in $X$. 

Firstly, since the disformal transformation amounts to the change of units of time and length as 
$d\tilt=\alpha dt, d\tilx=\A^{1/2}dx$,   eq.~(\ref{conserved}) now becomes
\beqa
C^2_E dS={\rm constant} \,, 
\label{conserved:e-frame}
\eeqa
where $C^2_E=g^{\mu\nu}C_{\mu}C_{\nu}=\A C^2$ and $dS=\A^{-1}d\tilS$.  
From eq.~(\ref{T_correspondence}), the energy-momentum tensor of light rays in 
the Einstein frame under the geometric optics approximation up to  the first order 
in $X$ is 
\begin{align}
    T_{\mu\nu} 
    &=  \alpha \A^{1/2} \left[\Ttilde_{\mu\nu} -
        \B\left(\phi_{\mu}\phi_{\alpha}{\Ttilde^{\alpha}}_{~\nu}+\phi_{\nu}\phi_{\alpha}{\Ttilde^{\alpha}}_{~\mu}\right)\right]
\nonumber \\
    &= \frac{\alpha}{\A^{1/2}}C^2_E
        \left[ k_{\mu}k_{\nu} - \frac{\B (k_{\alpha}\phi^{\alpha})}{\alpha^2} 
            \left(\phi_{\mu}k_{\nu}+\phi_{\nu}k_{\mu}\right) \right] \,.
\end{align}
Thus, the measured energy flux corresponding to eq.~(\ref{fluxvector})  becomes
\beqa
f^{\mu}=-h^{\mu\nu}T_{\nu\rho}u^{\rho}=f n^{\mu} \,,
\eeqa
where $n^{\mu}$ is the unit vector in the spatial direction of the propagation vector seen 
on the rest frame of the observer and $f $ is the observed flux that are given by 
\begin{subequations}
\beqa
h^{\mu\nu}&=&g^{\mu\nu}+u^{\mu}u^{\nu} \,,\\
k^{\mu}&=&(-u_{\nu}k^{\nu})\left(u^{\mu}+\frac{\A^{1/2}}{\alpha}n^{\mu}\right)=
\omega \left(u^{\mu}+\frac{\A^{1/2}}{\alpha}n^{\mu}\right) \,,\\
f&=&\frac{\A}{\alpha^2}C^2_E\omega^2 \,.
\label{flux:e-frame}
\eeqa
\end{subequations}
However, care must be taken in defining the luminosity of the source $L_S$ 
from the measured flux  because the photon number is no longer conserved in 
the Einstein frame. This is seen 
from eq.~(\ref{amplitude}) which implies the existence of the conserved current 
$C^2\tilk^{\mu}$ and the photon number density $C^2\tilomega$ in Jordan frame.  
On the other hand,  eq.~(\ref{amplitude}) implies that the conserved current in 
the Einstein frame is
\beqa
\frac{\alpha}{\A^{1/2}}C_E^2\left(k^{\mu}-\frac{\B( k_{\alpha}\phi^{\alpha})}{\alpha^2}\phi^{\mu}\right) \,,
\eeqa
and the density measured by the observer is $(\A^{1/2}/\alpha)C_E^2\omega$ and 
the spatial volume integral of it is conserved. 
This should be contrasted with the  energy density measured by the observer 
which is given by,  up to in the first order in $X$, 
\beqa
T_{\mu\nu}u^{\mu}u^{\nu}=\frac{\A^{3/2}}{\alpha^3}C_E^2\omega^2 \,.
\eeqa
Therefore, a bare number density $(\A^{3/2}/\alpha^3)C_E^2\omega=(\A/\alpha^2)\times(\A^{1/2}/\alpha)C_E^2\omega$ is not a conserved quantity and 
we need to use the ``modified number density''  $(\A^{1/2}/\alpha)C_E^2\omega$ and hence the modified flux $f_E$, given by
\beqa
f_E=(\alpha^2/\A)f=C^2_E\omega^2 \,,
\label{fe-f}
\eeqa
to relate the measured flux to the source's luminosity. From eq.~(\ref{conserved:e-frame}), 
this $f_E$ obeys the flux conservation law: $f_E\propto dS^{-1}$. 
Thus, we define the source's luminosity $L_S$ through the modified flux $f_{ES}$   measured  on a small sphere around the source with the radius $\Delta r$ as
\beqa
L_S=4\pi\Delta r^2 f_{ES} =4\pi \Delta r^2C_{ES}^2 \omega_S^2 \,. 
\eeqa
In terms of the source's luminosity $L_S$ measured in the observer's unit, 
the luminosity distance $d_L$ is defined as $d_L=\sqrt{L_S(\alpha_O/\alpha_S)^2/4\pi f_{EO}}$. 

We define the source area distance $r_S$ and the observer area distance $r_O$ by
\beqa
&&dS_S=r_S^2 d\Omega_S,\\
&&dS_O=r_O^2 d\Omega_O,
\eeqa
where $dS_S$ is the cross-sectional area at the observer of the flux 
from the source and $dS_O$ is the cross-sectional area of 
source object. 

The relation between $r_S$ and $d_L$ is found as follows: 
from eq.~(\ref{conserved:e-frame}),  
one can show that
$C_{ES}^2\Delta r^2d\Omega_S=C_{EO}^2 dS_S$. 
Then, using eq.~(\ref{redshift:einstein}),
\beqa
d_L^2=\frac{L_S(\alpha_O/\alpha_S)^2}{4\pi f_{EO}}=\frac{C_{ES}^2}{C_{EO}^2}
\frac{\omega_S^2}{\omega_O^2}\frac{\alpha_O^2}{\alpha_S^2}\Delta r^2=(1+z)^2\frac{dS_S}{d\Omega_S}=(1+z)^2r_S^2 \,.
\label{dl-rs:e-frame}
\eeqa
However,since the source's area $dS_O$ measured by the observer's unit is 
$dS_O(\A_S/\A_O)$, the angular diameter distance $d_A$ is related to $r_O$ as
\beqa
d_A^2=\frac{dS_O(\A_S/\A_O)}{d\Omega_O}=r_O^2\frac{\A_S}{\A_O}.
\label{da-ro:e-frame}
\eeqa

\begin{figure}[tbp]
	\centering
	\includegraphics[width=0.55\textwidth]{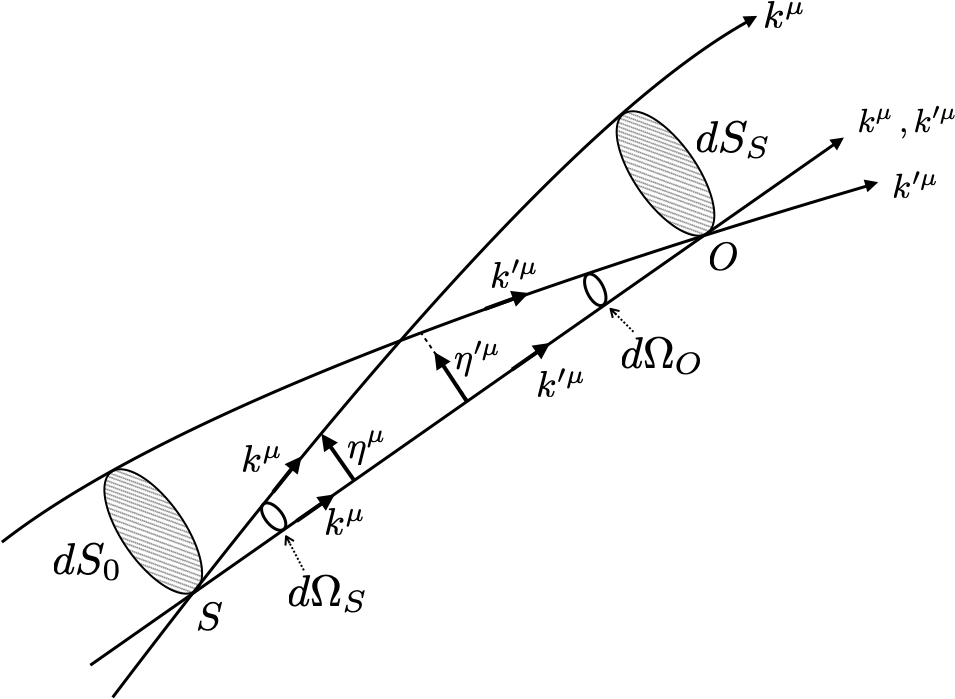}
	\caption{ \label{fig:light_bundle}
 	{Schematic picture of congruences of photons expanding from a  source $S$ 
 and converging to an observer $O$, with a common geodesic $SO$.} }
\end{figure}

Next, following~\cite{ellis}, we show that  the following reciprocity relation in the 
Einstein frame corresponding to eq.~(\ref{reciprocity}) holds 
although the geodesics deviation equation is modified :
\beqa
r_S=(1+z)r_O\left(\frac{\A_S}{\A_O}\right)^{1/2}.
\label{reciprocity:einstein}
\eeqa
Let a bundle of light rays diverging from $S$ with the solid angle $d\Omega_S$ 
have tangent vector $k^{\mu}$, and let a bundle of light rays converging to $O$ 
with the solid angle $d\Omega_O$ have tangent vector $k'_{\mu}$, where $OS$ is a 
geodesics common to both bundles so that $v=v'$ and $k^{\mu}=k'^{\mu}$  
on $OS$ (see figure~\ref{fig:light_bundle}).
Let $v, v'$ be affine parameters and $\eta^{\mu}, \eta'^{\mu}$ be the Jacobi vectors for $k^{\mu}, k'^{\mu}$, respectively. 
Then, $k^{\mu}={ \partial x^{\mu}/}{\partial v}$ 
and  from eq.~(\ref{geodesics:e-frame})
\begin{subequations} \label{jacobi}
\begin{align} 
\frac{Dk_{\mu}}{D v}
    &= k^{\nu} \nabla_{\nu} k_{\mu}
    = \frac{1}{2} \nabla_{\mu} \qty( \frac{\B \qty( \phi^{\mu} k_{\mu} )^2 }{ \A - 2\B X } )
    \equiv \nabla_{\mu} K \,,
\\
\frac{D\eta_{\mu}}{Dv}
    &= k^{\nu} \nabla_{\nu} \eta_{\mu}
    = \eta^{\nu} \nabla_{\nu} k_{\mu} \,,
\end{align} 
\end{subequations}
and similar equations for primed variables. From eq.~(\ref{jacobi}), we can derive a modified geodesics deviation equation
\beqa
\frac{D^2\eta_{\mu}}{Dv^2}=k^{\rho}\nabla_{\rho}(k^{\nu}\nabla_{\nu}\eta_{\mu})=
-R_{\mu\nu\rho\sigma}k^{\nu}\eta^{\rho}k^{\sigma}+\eta^{\nu}\nabla_{\nu}\nabla_{\mu}K,
\eeqa
and similarly for primed variables. Then along $OS$ we have 
\beqa
\eta'^{\mu}\frac{d^2\eta_{\mu}}{dv^2}-\eta^{\mu}\frac{d^2\eta'_{\mu}}{dv^2}=0,
\eeqa
which implies 
\beqa
\eta'^{\mu}\frac{d \eta_{\mu}}{dv}-\eta^{\mu}\frac{d\eta'_{\mu}}{dv}=0
\eeqa
along $OS$. 
Since $\eta^{\mu}=0 $ at $S$ and $\eta'^{\mu}=0$ at $O$, this gives
\beqa
\eta'^{\mu}\frac{d\eta_{\mu}}{dv}\Bigg{|}_S=-\eta^{\mu}\frac{d\eta'_{\mu}}{dv}\Bigg{|}_O.
\label{condition:jacobi}
\eeqa
We assume that $\eta^{\mu}$ is orthogonal to the observer's four-velocity at $O$ and 
$\eta'^{\mu}$ is orthogonal to the source's four-velocity at $S$. 
Then as given in eq.~(\ref{flux:e-frame}), $k^{\mu}$ and $k'^{\mu}$ are decomposed into the temporal direction and 
 the spatial direction in the rest frame of observer's four-velocity:  
 $k^{\mu}=\omega(u^{\mu}+(\A^{1/2}/\alpha)n^{\mu})$. Thus 
an infinitesimal affine distance $dv$ corresponds to  the proper distance 
$d\ell$ as
\beqa
|d\ell|=\frac{\omega \A^{1/2}}{\alpha}|dv|.
\label{ell}
\eeqa
We choose a pair of  Jacobi fields $\eta^{\mu}_1, \eta^{\mu}_2$ such that 
\beqa
\frac{d\eta^{\mu}_1}{dv}\frac{d\eta_{2\mu}}{dv}\Bigg{|}_S=\eta_1^{\mu}\eta_{2\mu}\Bigg{|}_O=0.
\eeqa 
This suggests we may take ${\eta'_1}^{\mu}|_S\propto \eta_1^{\mu}|_S$ and 
${\eta'_2}^{\mu}|_S\propto \eta_2^{\mu}|_S$. So, 
we  choose a pair of  Jacobi fields $\eta'^{\mu}_1, \eta'^{\mu}_2$ determined 
by the condition
\beqa
{\eta'_1}^{\mu}\frac{d\eta_{2\mu}}{dv}\Bigg{|}_S={\eta'_2}^{\mu}\frac{d\eta_{1\mu}}{dv}\Bigg{|}_S=0.
\eeqa
Then from eq.~(\ref{condition:jacobi}), we have
\beqa
\eta_1^{\mu}\frac{d\eta'_{2\mu}}{dv}\Bigg{|}_O=\eta_2^{\mu}\frac{d\eta'_{1\mu}}{dv}\Bigg{|}_O=0.
\eeqa
Hence, we have
\beqa
{\eta'_1}^{\mu}\eta'_{2\mu}\Bigg{|}_S=\frac{d\eta'^{\mu}_1}{dv}\frac{d\eta'_{2\mu}}{dv}\Bigg{|}_O=0.
\eeqa
With this choice of Jacobi vectors,
\beqa
dS_S=\eta_1\eta_2|_O,~~~dS_O=\eta'_1\eta'_2|_S,~~~
d\Omega_S=\frac{d\eta_1}{d\ell}\frac{d\eta_{2}}{d\ell}\Bigg{|}_S,~~~
d\Omega_O=\frac{d\eta'_1}{d\ell}\frac{d\eta'_{2}}{d\ell}\Bigg{|}_O,
\eeqa
where $\eta_1, \eta'_1$ are the norms of $\eta_1^{\mu},{\eta'_1}^{\mu}$.  On the other hand, from 
eq.~(\ref{condition:jacobi}), we have
\beqa
\eta_1\eta_2\frac{d\eta'_{1}}{dv}\frac{d\eta'_{2}}{dv}\Bigg{|}_O=
\eta'_1\eta'_2\frac{d\eta_{1}}{dv}\frac{d\eta_{2}}{dv}\Bigg{|}_S.
\eeqa
Therefore, using eq.~(\ref{ell}) and eq.~(\ref{redshift:einstein}), we finally obtain
\beqa
dS_Sd\Omega_O\tilomega_O^2\A_O=dS_Od\Omega_S\tilomega_S^2\A_S. 
\eeqa
In terms of the area distances $r_S$ and $r_O$, this gives eq.~(\ref{reciprocity:einstein}) : 
 $r_S=(1+z)r_O(\A_S/\A_O)^{1/2}$. 
Since $r_S$ is not a measurable quantity, this modification has no observable effect.  
However, in terms of observable distances $d_L$ and $d_A$ 
using eq.~(\ref{dl-rs:e-frame}) and eq.~(\ref{da-ro:e-frame}),  we find 
the distance-duality relation 
\beqa
d_L=(1+z)^2d_A. 
\eeqa
 Therefore, the distance-duality 
relation is invariant under the disformal transformation although 
the reciprocity relation is modified.

\subsection{Non-universal coupling}
\label{subsec:non-universal}

We have assumed that the disformal coupling is universal for all matter species. 
We here comment on the case of non-universal disformal coupling.
\footnote{Disformally coupled dark matter theories were also investigated in~\cite{Gleyzes:2015rua, vandeBruck:2016hpz, Mifsud:2017fsy, Kimura:2017fnq, Chibana:2019jrf}.}
For example, let us consider the case where we have dark matter frame metric 
$g_{M\mu\nu}$ in addition to  Jordan (photon)  frame metric $\gtilde_{\mu\nu}$ 
and  Einstein frame metric $g_{\mu\nu}$, namely
\beqa
g_{M\mu\nu}=\A_Mg_{\mu\nu}+\B_M\phi_{\mu}\phi_{\nu}
\eeqa
and eq.~(\ref{eq:metric_transf}) for $\gtilde_{\mu\nu}$. 
Even in this case,  the distance-duality relation is invariant under 
the disformal transformation from $\gtilde_{\mu\nu}$  to $g_{M\mu\nu}$
because $\gtilde_{\mu\nu}$ is related to $g_{M\mu\nu}$ as
\beqa
\gtilde_{\mu\nu}=\frac{\A}{\A_M}g_{M\mu\nu}+\left(\B-\frac{\A}{\A_M}\B_M\right)\phi_{\mu}\phi_{\nu}\,.
\eeqa
Hence, we can similarly define
\beqa
\alpha_M=\sqrt{\A/\A_M-(\B-\B_M\A/\A_M)(\phi_{\mu}u_M^{\mu})^2},
\eeqa
where $u_M^{\mu}=dx^{\mu}/d\tau_M$ with $d\tau_M^2 = - g_{M \mu\nu} dx^{\mu} dx^{\nu}$.
Therefore,  the conclusion concerning photons in section \ref{sec:cosmology} (and 
section \ref{subsec:distribution} and section \ref{subsec:specific} below) still holds for non-universal coupling case as long as $\alpha_M$ is well-defined.

\section{Perturbed observables}
\label{sec:perturbed_observables}

\subsection{Photon distribution function}
\label{subsec:distribution}

We consider observable quantities related to cosmic microwave background.  
Firstly, we study the frame-dependence of the photon distribution function. 

First of all, we note that in the geometric optics approximation, 
the photon four-momentum $p_{\mu}$, conjugate to the comoving coordinate $x^{\mu}$,
is defined as $p_{\mu}=\hbar k_{\mu}$, 
where $k_{\mu}=\partial_{\mu}S$ is the propagation vector, invariant under disformal transformations.
Therefore, the phase-space volume element $ dx^1dx^2dx^3dp_1dp_2dp_3=d^3xd^3p$ is 
frame-invariant. 

Moreover, in the geometric optic approximation, from eq.~(\ref{T_correspondence}), 
the relationship between the Einstein frame and Jordan frame  
energy-momentum tensor is given by
\beqa
T^{\mu\nu}=\alpha\A^{5/2}\tilT^{\mu\nu},
\eeqa
The energy-momentum tensor can be written as integrals over phase space and is,   respectively, given by 
\beqa
T^{\mu\nu}=\frac{2}{(2\pi)^3}\int\frac{d^3p}{\sqrt{-g}p^0}p^{\mu}p^{\nu}f  \,, \qquad
\tilT^{\mu\nu}=\frac{2}{(2\pi)^3}\int\frac{d^3p}{\sqrt{-\tilg}\tilp^0}\tilp^{\mu}\tilp^{\nu}\tilf \,,
\eeqa
where $p^{\mu}=dx^{\mu}/d\lambda$ and $\tilp^{\mu}=dx^{\mu}/d\tillambda$. Hence, the relation between Jordan frame and the Einstein frame distribution function is given by 
\cite{vandeBruck:2013yxa}
\beqa
\tilf=\frac{1}{\A}\frac{d\tillambda}{d\lambda}f.
\label{tilf-f1}
\eeqa
{}From the relation of the measured frequency,  
$\tilomega=-\tilk^{\mu}\tilu_{\mu}=\omega/\alpha=(-k^{\mu}u_{\mu})/\alpha$,  we find 
$d\tillambda=\alpha^2d\lambda$. Therefore,  we have
\beqa
\tilf=\frac{\alpha^2}{\A}f.
\label{tilf-f2}
\eeqa
Hence, the number of photons in the phase space in Jordan frame $d\tilN=\tilf d^3xd^3p$ 
is related to that in the Einstein frame $dN=fd^3xd^3p$ by $d\tilN=(\alpha^2/\A)dN$. 
This factor $\alpha^2/\A$ is consistent with 
the difference of the number density of photons between Jordan frame and the Einstein 
frame  discussed in subsection~\ref{sec:rec_relation}.

In Jordan frame, the rate of change of the distribution function along the photon world-line is given by the Boltzmann equation,
\begin{equation} \label{eq:boltzmann_jf}
    \frac{D\tilf}{D\tillambda}= \frac{d{x^{\mu}}}{d\tillambda} \pdv{\tilf}{{x^{\mu}}} +\frac{d{p_{\mu}}}{d\tillambda} \pdv{\tilf}{p_{\mu}} = \tilC[\tilf ] \,,
\end{equation}
where $\tilC$ is the collision term.  

In the case of the conformal transformation, since the electromagnetic Lagrangian, eq.~\eqref{action:photon}, is conformal-invariant, the Boltzmann equation is also invariant under conformal transformations~\cite{cy}. 
On the other hand, in the case of the disformal transformation, the distribution function  is no longer invariant and considering eq.~(\ref{tilf-f2}), 
 the Boltzmann equation is modified, 
\beqa
\frac{D f}{D\lambda}= \frac{d{x^{\mu}}}{d\lambda} \pdv{f}{x^{\mu}} +\frac{d{p_{\mu}}}{d{\lambda}} \pdv{f}{p_{\mu}} =\A\tilC[(\alpha^2/\A)f ]-\frac{\A}{\alpha^2}\left(\frac{D }{D\lambda}\left(\frac{\alpha^2}{\A}\right)\right)f.
\eeqa
It is to be noted that this does not immediately imply that the spectrum of temperature anisotropies is frame-dependent. The detailed studies of  
the temperature anisotropies are left for  future work. 


\subsection{Specific flux and specific intensity}
\label{subsec:specific}

Next, we study the specific flux and specific intensity of CMB photons.

Let us start with  Jordan frame, in which the discussion follows closely the standard general relativity case~\cite{ellis,Ellis:2013cu}.
The flux measured  in frequencies between $\tilomega$ and 
$\tilomega + d\tilomega$ is given by
\begin{equation} \label{eq:spec_flux_jf}
    \tilf_{\omega} d\tilomega 
        = \frac{\tilL_S}{4\pi} \frac{1}{\tilr_S^2(1+z)} \tilde{\I} \qty[ (1+z) \tilomega] d\tilomega ,
\end{equation}
where $\tilf_{\omega}$ is the specific flux and $\tilomega$ is the observed frequency 
and  we drop the subscript  $O$ to simplify the notation. 
The function $\tilde{\I}(\tilomega)$ represents the source spectrum, such that $\tilL_S \tilde{\I} d\tilomega$ is the source's rate of emission in the frequency range ($\tilomega, \tilomega + d\tilomega$).
Note that  $\tilomega$ is the observed frequency, corresponding to a frequency $(1+z) \tilomega$ at emission.
Then, the specific intensity $\tilI_{\omega}$,  the flux measured in a solid angle 
$d\Omega_O$ can be written as
\begin{equation} \label{eq:spec_int_jf}
    \tilI_{\omega} d\tilomega = \frac{\tilf_{\omega} d\tilomega}{d\Omega_O}
        =\frac{\tilL_S}{4\pi d\tilS_O}  \qty( \frac{\tilr_O}{\tilr_S} )^2 
        \frac{\tilde{\I} \qty[ (1+z) \tilomega ] d\tilomega}{(1+z)}
        =\tilI_S\frac{ \tilde{\I} \qty[ (1+z) \tilomega ] d\tilomega }{ (1+z)^3} ,
\end{equation}
where $\tilI_S = \tilL_S / 4\pi d\tilS_O$ is the surface brightness (an intrinsic property of the source),
and we have used the definition of observer area distance eq.~\eqref{eq:ro_jf} 
in the 
second equality and used the reciprocity relation eq.~\eqref{reciprocity} in the third equality. 
We see that the observed specific intensity does not depend on the area distance of the source.

Now let us consider a source emitting black-body radiation, such as the CMB spectrum, 
\begin{equation}
    \tilI_S \tilde{\I} (\tilomega_S) = \tilomega_S^3 \tilF \qty(\frac{\tilomega_S}{\tilT_S}) \,, \qquad
    \tilF \qty(\frac{\tilomega}{\tilT}) = \frac{4\pi \hbar}{c^2} \frac{1}{e^{\hbar\tilomega/k\tilT} - 1},
\end{equation}
where $\tilF (\tilomega/\tilT)$ is the Planck function for black-body radiation with temperature $\tilT$ and we have restored 
$\hbar$ and $c$. 
Then, eqs.~\eqref{eq:spec_int_jf} and~\eqref{eq:ro_jf} lead to
\begin{equation}
    \tilI_{\omega} = \tilomega^3 \tilF \qty(\frac{\tilomega}{\tilT(z)})
\end{equation}
where $\tilT(z) = \tilT_S/(1+z)$ is the observed temperature, redshifted due to 
the expansion of the universe, and the measured intensity does not deviate from the black-body spectrum.

In Einstein frame, from eq.~\eqref{dl-rs:e-frame}, the observed specific flux, calculated from the conserved flux $f_E$, is given by
\begin{equation} \label{eq:spec_flux_ef}
    f_{\omega} d\omega 
        = \frac{\alpha_O}{\alpha_S} \frac{L_S}{4\pi} \frac{1}{r_S^2(1+z)} 
        \I\qty[\frac{\alpha_S}{\alpha_O} (1+z) \omega] d\omega \,,
\end{equation}
and, accordingly, the specific intensity, 
$I_\omega d\omega = f_\omega d\omega / d\Omega_O$,
becomes
\begin{equation}
    I_{\omega}d\omega
        = \frac{\alpha_O \A_O}{\alpha_S \A_S} \frac{1}{(1+z)^3}
            I_S \I \qty[ \frac{\alpha_S}{\alpha_O} (1+z) \omega ] d\omega \,,
\end{equation}
where $I_S = L_S / 4\pi dS_O$ is the surface brightness
and we have used the definition of observer area distance in Einstein frame, eq.~\eqref{eq:ro_jf}, 
and the reciprocity relation, eq.~\eqref{reciprocity:einstein}.
Once again, we consider the emission by a black-body source. 
Considering eq.~(\ref{tilf-f2}), the source intensity is given by
\begin{equation}
    I_S \I (\omega_S) = \omega_S^3 F_E\qty(\frac{\omega_S}{T_S}) \,, \qquad
    F_{E}\qty(\frac{\omega}{T}) = \frac{4\pi \hbar}{c_E^2} \frac{1}{e^{\hbar\omega/kT} - 1} \,,
\end{equation}
where $c_E^2=\alpha^2/\A$ is the effective speed of light squared, eq.~\eqref{speedoflight}.  
By taking into account the difference between $c_E$ at  the source and that at the observer,  
the measured intensity   $I_{\omega}$ is given by
\begin{equation}
    I_{\omega} = \omega^3 F_E\qty(\frac{\omega}{T(z)})\,,
\end{equation}
where, we have defined the observed temperature 
\begin{equation}
    T(z) = \frac{\alpha_O}{\alpha_S} \frac{T_S}{(1+z)}
\end{equation}
which takes into account both the expansion of the universe and the change of units from the source's reference frame to the observer frame, in accord with eq.~\eqref{redshift:einstein}. 
We conclude that $I_\omega$ is  given by the spectrum of black-body radiation with the temperature $T(z)$ and consequently there is no spectral distortion.

\subsection{Adiabaticity condition and non-adiabatic pressure perturbations}

Let us now consider the adiabaticity condition and non-adiabatic pressure perturbations.
We first derive the adiabaticity condition for a generic disformal transformation, and later, by choosing the functions $\A$ and $\B$, we discuss the particular cases.

The pressure perturbation of matter fields $\delta \tilP$ can be split into two parts~\cite{Bardeen:1980kt, Kodama:1985bj, Wands:2000dp}, 
\begin{equation} \label{eq:delta_p_jordan}
	\delta \tilP = \tilc_s^2 \delta \tilrho + \frac{d \tilP}{d \tilt} \tilGamma \,,
\end{equation}
where the first term, proportional to the perturbation of the energy density $\delta \tilrho$, corresponds to the adiabatic part, and the second to the entropic (non-adiabatic) one. $\tilt$ is the cosmic time in the Jordan frame defined in eq.~(\ref{tau:scale}).
The sound speed, $\tilc_s^2$, is defined as
\begin{equation} \label{eq:cs_jordan}
	\tilc_s^2 = \frac{d \tilP / d \tilt}{d \tilrho / d \tilt} \,,
\end{equation}
whereas $\tilGamma$ is the gauge-invariant non-adiabatic pressure perturbation, which measures the displacement between hypersurfaces of uniform (adiabatic) pressure and uniform energy density.
In terms of the Einstein frame quantities, $\delta \rho$ and $c_s^2$ are similarly defined using the corresponding variables without the tilde mark:
\begin{align}
\label{eq:delta_P_einstein}
	\delta P &= c_s^2 \delta \rho + \frac{d P}{dt} \Gamma \,, \\
\label{eq:cs_einstein}
	c_s^2 &= \frac{dP / dt}{d \rho / dt} \,.
\end{align}

Before computing the relation between the entropy perturbations $\tilGamma$ and $\Gamma$, let us further discuss the relationships between the energy densities and pressures in Einstein and Jordan frames (see also appendix~\ref{sec:emt}).
To get the components of the energy-momentum tensor, we must first split $T_{\mu\nu}$, as measured by an observer with four-velocity $u^\mu$, into parts parallel and orthogonal to $u^\mu$, 
\begin{equation}
    T_{\mu\nu} = 
        \rho u_\mu u_\nu + q_\mu u_\nu + q_\nu u_\mu + P h_{\mu\nu} + \pi_{\mu\nu} \,,
\end{equation}
where, its 10 components are
\begin{align}
    \rho &= T_{\mu\nu} u^\mu u^\nu \,, \\
    P &= \frac{1}{3} h^{\mu\nu} T_{\mu\nu} \,, \\
    q_\mu &= - {h^\nu}_\mu T_{\nu\rho} u^\rho \,, \\
    \pi_{\mu\nu} &= {h^\rho}_\mu {h^\sigma}_\nu T_{\rho \sigma} - \frac{1}{3}  h^{\rho\sigma} T_{\rho\sigma} h_{\mu\nu}\,, 
\end{align}
the energy density, pressure, momentum density and anisotropic stress tensor, respectively, and 
\begin{equation}
    h_{\mu\nu} = g_{\mu\nu} + u_\mu u_\nu 
\end{equation}
is the spatial metric relative to the observer. 
The decomposition of $\tilT_{\mu\nu}$ is done in a similar way, with respect to tilded variables.
Moreover, to determine the relationship between the respective Jordan-frame and Einstein-frame components, we need the relation between the four-velocities of the observers in each frame, namely, $\tilu^\mu = u^\mu / \alpha$ as defined in eq.~\eqref{eq:four_velocity}, where $\alpha=\sqrt{\A - \B(\phi_\mu u^\mu)^2}$.

Since we are interested in cosmological applications, we consider linear perturbations around a homogeneous and isotropic background in both frames such that $q_\mu$, $\pi_{\mu\nu}$, $\tilde{q}_\mu$, and $\tilde{\pi}_{\mu\nu}$ are first-order quantities.
%
%
Furthermore, we take a comoving Jordan-frame observer, that is, $\tilu^i = 0$. According to eq.~\eqref{eq:four_velocity}, this choice implies $u^i = 0$ such that the equality $(\phi_\mu u^\mu)^2 = 2X$ is valid up to the linear order.
\footnote{We could also have chosen a Jordan-frame observer comoving with the fluid, such that $\tilu^\mu$ is an eigenvector of $\tilT^{\mu\nu}$ and $\tilde{q}_\mu$ vanishes. However, the corresponding Einstein-frame four-velocity, given by eq.~\eqref{eq:four_velocity}, would not be an eigenvector of $T^{\mu\nu}$.
This can be seen from the transformations~\eqref{eq:emt_equiv}, which would still be valid to the linear order. In particular, eq.~\eqref{eq:q_equiv2} shows that $\tilde{q}_\mu = 0$ does not necessarily implies that $q_\mu$ vanishes.}

The components of the energy-momentum tensor measured in Jordan frame are then related to the ones in Einstein frame as 
\begin{subequations} \label{eq:emt_equiv}
\begin{align}
\label{eq:rho_equiv2}
	\tilrho &= \xi_1 \rho + \xi_2 P \,,
\\
\label{eq:p_equiv2}
	\tilP &= \frac{P}{\alpha \A^{3/2}} \,,
\\
\label{eq:q_equiv2}
    \tilde{q}_\mu &= \frac{1}{\A^{3/2}} 
        \qty[ q_\mu - \zeta \phi_\rho u^\rho {h^\nu}_\mu \phi_\nu] \,,
\\
    \tilde{\pi}_{\mu\nu} &= \frac{\pi_{\mu\nu} }{ \alpha \A^{1/2}} \,, 
\end{align}
\end{subequations}
where, for convenience, we have defined the functions
\begin{align}
    \xi_1 &= {\frac
			{\left( 1 - 2\B X / \A \right)^{1/2}}
			{\A \left( \A - \A_{,X}X + {2} \B_{,X} X^2\right)} } \,, 
\\
    \xi_2 &= - {\frac
			{3 \left( 1 - 2\B X / \A \right)^{1/2} \A_{,X} X}
			{\A^2 \left( \A - \A_{,X}X + {2} \B_{,X} X^2\right)} } \,,
\\
    \zeta &= \frac{\B P}{\A - 2\B X} 
        + \frac{(\A_{,X} - 2\B_{,X} X) \rho - 3 \A_{,X} P }{2\left( \A - \A_{,X}X + {2} \B_{,X} X^2\right)} \,.
\end{align}
We once again point out that the above equations are valid up to linear order.
Note that, if we allow the function $\A$ to depend on $X$, the energy density in Jordan frame will explicitly depend, not only on the Einstein-frame $\rho$, but also on its pressure $P$.
This is a reflex of the fact that if $\A = \A(X)$, the metric transformation, eq.~\eqref{eq:metric_transf}, is non-linear in $g_{\mu\nu}$, since $X$ carries a hidden metric contraction.

A small comment is in order.
We could also have chosen a Jordan-frame observer comoving with the fluid, such that $\tilu^\mu$ is an eigenvector of $\tilT^{\mu\nu}$ and $\tilde{q}_\mu$ vanishes. \
However, the corresponding Einstein-frame four-velocity, given by eq.~\eqref{eq:four_velocity}, would not be an eigenvector of $T^{\mu\nu}$.
This can be seen from the transformations~\eqref{eq:emt_equiv}, which would still be valid to the linear order. In particular, eq.~\eqref{eq:q_equiv2} shows that $\tilde{q}_\mu = 0$ does not necessarily implies that $q_\mu$ vanishes.
Nevertheless, it is still possible to choose observers comoving with the fluid in both Jordan and Einstein frames.
In that case, $\tilde{q}^\mu = q^\mu = 0$, but the four-velocities will have a small spatial perturbation $\tilde{v}^i$ and $v^i$. These are related by
\begin{equation}
    \tilu^0 = \frac{u^0}{\alpha}
\,, \qquad
    \tilde{v}^i
        = \alpha^2 \qty[ \frac{(\rho + P) v^i - \zeta \, \phid \, \delta \phi^i}{\alpha^2 \gamma + \A P} ]  
\,, \qquad
    \gamma = \frac{\A \rho - 3 \A_{,X} X P}{\A - \A_{,X} X + 2 \B_{,X} X^2} \,.
\end{equation}
Still, the transformations of the energy density, pressure, and anisotropic stress in eq.~\eqref{eq:emt_equiv} remain valid (at least up to the linear level).

The relation between the energy density and pressure perturbation in Jordan and Einstein frames can be readily computed from eqs.~\eqref{eq:rho_equiv2} and~\eqref{eq:p_equiv2}, 
\begin{align}
\label{eq:delta_rho_equiv}
	\delta \tilrho &= \bar{\xi}_1 \delta\rho + \bar{\rho} \delta\xi_1 + \bar{\xi}_2 \delta P + \bar{P} \delta\xi_2 \,, 
\\
\label{eq:delta_p_equiv}
	\delta \tilP &= \frac{\bar{P}}{\bar{\alpha} \bar{\A}^{3/2}} \left( \frac{\delta P}{\bar{P}} 
	    - \frac{\delta\alpha}{\bar{\alpha}} - \frac{3}{2} \frac{\delta \A}{\bar{\A}} \right) \,,
\end{align}
where we explicitly discriminate with bars the background quantities. In what follows, we drop this notation.
Also, note that since $\delta \A$, $\delta \alpha$, $\delta \xi_1$, and $\delta \xi_2$ depend on $X$, 
the perturbations $\delta \tilrho$ and $\delta \tilP$ implicitly include the scalar perturbations to $g_{\mu\nu}$, the Einstein frame metric.

Now, we turn our attention to the non-adiabatic pressure perturbations. 
First, we rewrite eq.~\eqref{eq:delta_p_jordan} as
\begin{equation}
	\tilGamma = \frac{\delta \tilP}{d\tilP / d\tilt} - \frac{\delta \tilrho}{d\tilrho / d\tilt} \,.
\end{equation}
Using the equations that describe the transformations of $\delta \tilrho$ and $\delta \tilP$, eqs.~\eqref{eq:delta_rho_equiv} and~\eqref{eq:delta_p_equiv}, respectively, along with the derivatives with respect to the time $\tilt$ of the energy density, eq.~\eqref{eq:rho_equiv2}, and pressure, eq.~\eqref{eq:p_equiv2}, the expression above can be recast as
\begin{align} \label{eq:Gamma_equiv}
\frac{d \ln{\tilP}}{d \tilt} \, \tilGamma
	&= \left( 1 - w \frac{\tilc_s^2}{\tilw} \frac{\rho}{\tilrho} \, \xi_2 \right) \frac{c_s^2}{w} \frac{d \ln{\rho}}{d t} \Gamma
\nonumber \\
    &\phantom{=(}
	+ \left[ \frac{c_s^2}{w} - \frac{\tilc_s^2}{\tilw} \frac{\rho}{\tilrho}
		\left( \xi_1 + \xi_2 c_s^2 \right) \right] \frac{\delta \rho}{\rho}
	- \frac{\tilc_s^2}{\tilw} \frac{\rho}{\tilrho} \left( \delta \xi_1 + w \delta \xi_2 \right)
    - \frac{\delta\alpha}{\alpha} - \frac{3}{2} \frac{\delta \A}{\A} \,,
\end{align}
where, using the definitions of the sound speeds on Jordan and Einstein frames, eqs.~\eqref{eq:cs_jordan} and~\eqref{eq:cs_einstein}, respectively, one can show that 
\begin{equation}
	\frac{\tilc_s^2}{\tilw \tilrho} = \frac{1}{\rho}
		{\frac
		{\left( \frac{c_s^2}{w} \frac{\rhod}{\rho} 
		    - \frac{\dot{\alpha}}{\alpha} - \frac{3}{2} \frac{\dot{\A}}{\A}  \right) }
		{\left[ \xi_1 \left(\frac{\rhod}{\rho} + \frac{\dot{\xi}_1}{\xi_1} \right) 
			  + w \xi_2 \left(\frac{c_s^2}{w} \frac{\rhod}{\rho} + \frac{\dot{\xi}_2}{\xi_2} \right) \right] }
		} \,.
\end{equation}
In the equation above, the dots over Einstein frame quantities on the right-hand side denote the derivative with respect to the time~$t$.
Now, suppose that the adiabaticity condition, that is, $\tilGamma = 0$, holds in Jordan frame. 
Then, according to eq.~\eqref{eq:Gamma_equiv}, it will also hold in Einstein frame ($\Gamma  = 0$) if
\begin{equation} \label{eq:adiabatic_cond}
\left[ \frac{c_s^2}{w} - \frac{\tilc_s^2}{\tilw} \frac{\rho}{\tilrho}
		\left( \xi_1 + \xi_2 c_s^2 \right) \right] \frac{\delta \rho}{\rho}
	= \frac{\tilc_s^2}{\tilw} \frac{\rho}{\tilrho} \left( \delta \xi_1 + w \delta \xi_2 \right)
	+ \frac{\delta\alpha}{\alpha} + \frac{3}{2} \frac{\delta \A}{\A} \,.
\end{equation}
This expression represents the adiabaticity condition for a generic disformal transformation with $\A(\phi,X)$ and $\B(\phi,X)$.

In what follows, we make a simplification, assuming both $\A$ and $\B$ depend only on the scalar field, $\A = \A(\phi)$ and $\B = \B(\phi)$, and not on the kinetic term $X$.
With this assumption, the energy density in Jordan frame, eq.~\eqref{eq:rho_equiv2}, will depend on $\rho$, without any explicit dependence on the Einstein frame pressure $P$, and $\xi_1 = \alpha \A^{-5/2}$.
In this case, eq.~\eqref{eq:adiabatic_cond} becomes
\begin{align} \label{eq:adiabatic_cond2}
    \qty[ \qty( 1 + \frac{c_s^2}{w} ) \frac{\dot{\alpha}}{\alpha}
        + \qty( \frac{3}{2} - \frac{5}{2} \frac{c_s^2}{w} ) \frac{\dot{\A}}{\A} ] \frac{\delta\rho}{\rho}
    &= \qty[ \qty( 1 + \frac{c_s^2}{w} ) \frac{\rhod}{\rho} - 4 \frac{\dot{\A}}{\A} ] \frac{\delta\alpha}{\alpha} 
\nonumber \\
    &\phantom{=[}
        + \qty[ \qty( \frac{3}{2} - \frac{5}{2} \frac{c_s^2}{w} ) \frac{\rhod}{\rho} + 4 \frac{\dot{\alpha}}{\alpha} ] \frac{\delta\A}{\A} \,.
\end{align}
To proceed with the discussion, we identify two particular cases in which we fix either $\A$ or $\B$.

\paragraph{Conformal case:}
for a conformal transformation $\A = \A(\phi)$ and $\B = 0$, such that 
$\alpha^2 = \A$, the condition eq.~\eqref{eq:adiabatic_cond2} translates into
\begin{equation}
	\left( 1 - \frac{c_s^2}{w} \right) 
	\left( \frac{\dot{\alpha}}{\alpha} \frac{\delta \rho}{\rho}  - \frac{\dot{\rho}}{\rho}\frac{\delta \alpha}{\alpha} \right)
	= 0 \,.
\end{equation}
This equation has two solutions~\cite{cy},
\begin{equation} \label{eq:adiabatic_cod_conf}
	c_s^2 = w
	\,, \text{ or } \,
	\frac{\delta \rho}{\dot{\rho}}
		= \frac{\delta \alpha}{\dot{\alpha}} 
		= \frac{\delta\phi}{\dot{\phi}} \,,
\end{equation}
the first corresponding to $w = $ constant,
whereas the second, to the situation in which the entropy perturbation between the matter field and the scalar field $\phi$ vanishes.

\paragraph{Purely disformal case:}
on the other hand, for a pure disformal transformation $\A = 1$ and $\B = \B(\phi, X)$, one finds
$\alpha^2 = 1 - 2 \B X$, as well as $\xi_1=\alpha/(1+ {2} \B_{,X} X^2)$ and $\xi_2 = 0$, 
along with the adiabaticity condition
\begin{equation} \label{eq:adiabatic_cond3}
    \qty( \frac{\dot{\alpha}}{\alpha}
        + \frac{c_s^2}{w} \frac{\dot{\xi_1}}{\xi_1} ) \frac{\delta\rho}{\rho}
    = \qty( \frac{\rhod}{\rho} + \frac{\dot{\xi_1}}{\xi_1} ) \frac{\delta\alpha}{\alpha}
        + \qty( \frac{c_s^2}{w} \frac{\rhod}{\rho} - \frac{\dot{\alpha}}{\alpha} ) \frac{\delta\xi_1}{\xi_1} \,,
\end{equation}
where
\begin{equation}
  \frac{\dot{\xi_1}}{\xi_1} = \frac{\dot{\alpha}}{\alpha} {-} \frac{ {2} d(\B_{,X} X^2)/dt}{1+\B_{,X} X^2} \,.   
\end{equation}
For $\A = 1$ and $\B = \B(\phi)$, this adiabaticity condition reduces to
\begin{equation}
	\left( 1 + \frac{c_s^2}{w} \right) 
	\left( \frac{\dot{\alpha}}{\alpha} \frac{\delta \rho}{\rho}  - \frac{\dot{\rho}}{\rho}\frac{\delta \alpha}{\alpha} \right)
	= 0 \,,
\end{equation}
which, similarly to the conformal case, also translates into two possibilities:
\begin{equation} \label{eq:adiabatic_cond_disf}
	c_s^2 = -w 	
	\,, \text{ or } \,
	\frac{\delta \rho}{\dot{\rho}}
		= \frac{\delta \alpha}{\dot{\alpha}} 
		= \frac{\B_{,\phi} X \delta\phi + \B \delta X}{ \B_{,\phi} X \phid + \B \dot{X}} \,.
\end{equation}
The first possibility, $c_s^2=dp/d\rho=-w=-p/\rho$,  corresponds 
to the equation of state $P \propto \rho^{-1}$,
which characterizes the Chapligyn gas~\cite{Kamenshchik:2001cp}.
According to eqs.~\eqref{eq:rho_equiv2} and~\eqref{eq:p_equiv2}, 
for $\A = 1$ and $\B = \B(\phi)$, one has $\xi_1 = \alpha$ and $\xi_2 = 0$, which leads to the invariance of 
the quantity $\tilrho\tilP = \rho P$ under disformal transformations. 
Therefore, a Chaplygin gas in Jordan frame still behaves as a Chaplygin gas in Einstein frame,
such that the adiabaticity condition is not violated.
On the other hand, the second possibility, when compared to the conformal case, now also has a $X$ dependence coming from the disformal part.
To ensure that the entropy perturbation between matter and the scalar field vanishes, one must also take into account the perturbations to the metric, hidden inside $\delta X$.

In conclusion, the notion of adiabaticity is generally not invariant under disformal transformations.

\subsubsection{Comments on curvature perturbations}

In ref.~\cite{cy}, it was shown that the uniform-density curvature perturbation $\tilde{\zeta}$ is conformally invariant when the adiabaticity condition eq.~\eqref{eq:adiabatic_cod_conf} is satisfied (in the case of constant equation of state, $w$ has to be $1/3$).
In  Jordan frame, the conservation of energy and momentum leads to the conservation of the uniform-density curvature perturbation on super-horizon scales when the non-adiabatic pressure perturbation $\Gamma$ is negligible~\cite{Wands:2000dp}.
However, since the matter energy-momentum tensor in the Einstein frame is not conserved, the condition $\Gamma = 0$ by itself does imply the conservation of $\zeta$ on large scales.
Despite that, since $\tilGamma = 0$ leads to the conservation of $\tilde{\zeta}$ on super-horizon scales, the adiabaticity condition ensures that $\zeta$ is also conserved on large enough scales.
This relation between the adiabaticity and invariance of the uniform-density curvature perturbation cannot be easily extended to the disformal case, not even in the simple case $\A = 1$, $\B = \B(\phi)$, illustrated in eq.~\eqref{eq:adiabatic_cond_disf} (the connection between the adiabaticity condition and invariance of curvature perturbations relies on the behavior of $\delta \A$).
In fact, in ref.~\cite{Minamitsuji:2014waa}, it was explicitly shown that $\Gamma = 0$ does not lead to the conservation of $\zeta$ on super-horizon scales for a disformal transformation with $\A = \A(\phi)$ and $\B = \B(\phi)$.

In the absence of matter ($\tilT_M^{\mu\nu} = T_M^{\mu \nu} = 0$), the comoving curvature perturbation $\widetilde{\mathcal{R}}$
is, not only conformally invariant~\cite{Makino:1991sg, Komatsu:1999mt, Chiba:2008ia}, 
but also disformally invariant under the transformation 
$\tilg_{\mu\nu} = \A(\phi)g_{\mu\nu} + \B(\phi) \phi_\mu\phi_\nu$~\cite{Minamitsuji:2014waa}.
On the other hand, for general disformal transformations of the type $\A(\phi, X)$, $\B(\phi, X)$, that is no longer the case, and the difference between the comoving curvature perturbation in each frame can be written in terms of the perturbation of a scalar field~\cite{Motohashi:2015pra}.
However, it can be shown that this quantity vanishes on super-horizon scales at least in the framework of the Horndeski theory such that the difference $\tilde{\mathcal{R}} - \mathcal{R}$ also vanishes~\cite{Motohashi:2015pra}.

\section{Summary}

In this work, we discussed the frame dependence and independence of cosmological observables under disformal transformations and provided the correspondence between several quantities in Jordan and Einstein frames.  

We have derived the correspondence between the geodesic equations for a test particle and photons, as well as the effective gravitational constant.  
We have found that the modification of the effective gravitational constant is the same as in the conformal case. 
We have also derived the geodesic equation in Einstein frame under the geometric optics approximation and have found that the tangent vector of the geodesics is no longer null.

We also investigated the redshift, the luminosity and angular diameter distances, and the observed specific intensity, finding that the redshift and the distance-duality relation are frame-independent and that the cosmic microwave background with initially Planck spectrum remains a Planck spectrum at the point of observation. 
In particular, we proved that, under the geometric optics approximation,  the distance-duality relation holds for general spacetime geometries, even though the reciprocity relation is modified.

Generally speaking, dimensionful quantities (such as the Hubble parameter, the gravitational constant, and the distances) are frame-dependent as in the case of the conformal transformation. 
Notably, the relation between the redshift drift and the acceleration of the universe is frame-dependent, which is also the case for the solution to the horizon problem.
Moreover, unlike the case of conformal transformations, the distribution function of photons is frame-dependent which stems from the fact that the photon number is no longer conserved.  
It would be interesting to study the temperature anisotropies in both Einstein  and  Jordan frames.

We have also investigated the adiabaticity conditions and non-adiabatic pressure perturbations, extending the discussions presented in ref.~\cite{Minamitsuji:2014waa} to a more generic case, and found that, in general, the adiabaticity condition is not invariant under disformal transformations.

In this paper, we confine ourselves to classical theory. It would be interesting to extend our arguments to quantum theory as done in refs.~\cite{Kamenshchik:2014waa,Ruf:2017xon,Ohta:2017trn} for the conformal case.

\section*{Acknowledgments}

We would like to thank Luca Buoninfante, Saikat Chakraborty, Atsushi Naruko for useful discussions. 
TC is supported  by MEXT KAKENHI Grant
Number 15H05894 and in part by Nihon University. 
FC and MY are supported in part by JSPS Bilateral Open Partnership Joint Research Projects.
MY is supported in part by JSPS Grant-in-Aid for Scientific Research Numbers 18K18764, MEXT KAKENHI Grant-in-Aid for Scientific Research on Innovative Areas Numbers 15H05888, 18H04579, Mitsubishi Foundation, JSPS and NRF under the Japan-Korea Basic Scientific Cooperation Program.

\appendix

\section{ Energy-momentum tensor}
\label{sec:emt}

In this appendix, we derive some relations regarding the matter energy-momentum tensor.

The energy-momentum tensor in Jordan frame, given by
\begin{equation}
	\Ttilde^{\mu \nu} = \frac{2}{\sqrt{-\gtilde}} \frac{\delta S_{M}}{\delta \gtilde_{\mu\nu}} \,,
\end{equation}
is covariantly conserved since matter is minimally coupled to Jordan frame metric $\gtilde_{\mu\nu}$, that is, 
\begin{equation}
	\nablatilde_{\mu} \Ttilde^{\mu\nu} = 0 \,.
\end{equation}
Furthermore, the energy-momentum tensor in the Einstein frame, $T^{\mu\nu}$, is related to $\Ttilde^{\mu \nu}$ as
\begin{equation}
	T^{\mu \nu} = \frac{2}{\sqrt{-g}} \frac{\delta S_{M}}{\delta g_{\mu\nu}}
	    =\sqrt{\frac{-\gtilde}{-g}} \frac{\delta \gtilde_{\alpha\beta}}{\delta g_{\mu\nu}}\Ttilde^{\alpha \beta}\,,
\end{equation}
and hence, it is not covariantly conserved: $\nabla_\mu T^{\mu \nu} \neq 0$.  
Using eq.~(\ref{eq:metric_transf}) and eq.~(\ref{determinant}) we can now write down the relation between the energy-momentum tensors in the two frames :
\begin{equation} \label{T_correspondence}
    T^{\mu\nu} = \A^{3/2}\left(\A-2{\B}X\right)^{1/2}
        \qty[ \A \Ttilde^{\mu\nu} + \frac{1}{2} \Ttilde^{\alpha\beta}
            \qty( \A_{,X} g_{\alpha\beta} + \B_{,X} \phi_{\alpha} \phi_{\beta} ) \phi^{\mu} \phi^{\nu} ] \,,
\end{equation}
which coincides with the result obtained in~\cite{Zumalacarregui:2012us} for a simpler 
case of $\A=\A(\phi),\B=\B(\phi)$. 
Written in another form, separating Jordan-frame and Einstein-frame variables,
\begin{equation} \label{T_correspondence2}
    \Ttilde^{\mu\nu} = \A^{-5/2} \qty( \A- 2\B X )^{-1/2}
        \qty[ T^{\mu\nu} 
            - \frac{\A_{,X} g_{\alpha\beta} + \B_{,X} \phi_{\alpha} \phi_{\beta} }{ 2 \qty( \A - \A_{,X} X + 2\B_{,X} X^2 ) }
            T^{\alpha\beta} \phi^{\mu} \phi^{\nu} ]\,.
\end{equation}

Let us now assume that matter in Jordan frame can be described by a perfect fluid, with the energy-momentum tensor, as measured by an observer with four-velocity $\tilde{u}^{\mu} = dx^{\mu}/d\tilde{\tau}$, being given by
\begin{equation}
\Ttilde^{\mu\nu}=(\tilde{\rho}+\tilde{P})\tilde{u}^{\mu}\tilde{u}^{\nu}+\tilde{P}\gtilde^{\mu\nu}.
\end{equation}
The four-velocities in Jordan and Einstein frame are related as
\begin{equation}
\tilde{u}^{\mu}=\frac{dx^{\mu}}{d\tilde{\tau}}=u^{\mu}\frac{d\tau}{d\tilde{\tau}} \,,
\end{equation}
such that, from eq.~\eqref{T_correspondence}, the energy-momentum tensor in the Einstein frame can be expressed as 
\begin{align}
T^{\mu\nu} 
    &= \A^{3/2} \qty( \A - 2 \B X )^{1/2}
        \Biggl[ \A \qty( \frac{d\tau}{d\tilde\tau} )^2 \qty( \tilde{\rho} + \tilde{P} ) u^{\mu} u^{\nu} 
            + \tilde{P} g^{\mu\nu} 
\nonumber \\
     &\phantom{= \A^{3/2} \qty( \A - 2 \B X )^{1/2} \Biggl[ }
        + \Biggl\{ 
                \frac{\tilde{P}}{\A-2\B X} 
                \qty[ \qty( 2\A - 3\B X ) \frac{\A_{,X}}{A} - \B - \B_{,X}X ] 
\\
    &\phantom{= \A^{3/2} \qty( \A - 2 \B X )^{1/2} \Biggl[ + \Biggl( }
          - \frac{1}{2} \qty( \frac{d\tau}{d\tilde\tau} )^2 \qty( \tilde{\rho} + \tilde{P} )
	      \qty[ \A_{,X} - \B_{,X} \qty( \frac{d\phi}{d\tau} )^2 ] \Biggr\}
	      \phi^{\mu}\phi^{\nu}  \Biggr] \,. 
\nonumber
\end{align}
From this equation, one can easily see that $T^{\mu\nu}$ does not necessarily take the form of a perfect fluid: it does so only when $u^{\mu} \propto \phi_{,}^{\mu}$, such as is the case of a homogeneous and isotropic universe.

Now, assume an FLRW metric in Einstein frame. 
The Jordan-frame metric is obtained by applying the disformal transformation,
\begin{eqnarray}
d\tilde{s}^2&=&\gtilde_{\mu\nu}dx^{\mu}dx^{\nu}\nonumber\\
&=&(\A g_{\mu\nu} + \B \phi_{\mu}\phi_{\nu})dx^{\mu}dx^{\nu}\nonumber\\
&=&-(\A - 2\B X)dt^2 + \A a^2(t){\gamma_{ij}}dx^i dx^j\,,
\end{eqnarray}
{where $\gamma_{ij}$ is the spatially homogeneous and isotropic
metric. It should be noticed that $X = \dot{\phi}(t)^2/2$, $\A =
\A(\phi(t),X(t))$, and $\B = \B(\phi(t),X(t))$ represent the background values.}
Since the Einstein-frame metric is an FLRW one, the four-velocity of a comoving observer in the Einstein frame is $u^{\mu}\equiv(1,\bar{0})$. Therefore the four-velocity of the comoving observer in  Jordan frame is
\begin{equation}
\tilde{u}^{\mu}=\frac{u^{\mu}}{(\A -2 \B X)^{1/2}}=\frac{1}{(\A -2\B X)^{1/2}}(1,\bar{0}) \,,
\end{equation} 
and the components of the energy-momentum tensor in Jordan frame become
\begin{equation}
    \Ttilde^{00} = \frac{\tilde{\rho}}{\A-2\B X} \,, \quad
    \Ttilde^{0i} = \Ttilde^{i0} = 0 \,, \quad
    \Ttilde^{ij} = \frac{\tilde{P}}{A a^2}{\gamma^{ij}} \,.
\end{equation}

Using eq.~(\ref{T_correspondence}), we can  calculate the components of the energy-momentum tensor in the Einstein frame. 
We find that the energy-momentum tensor remains diagonal in the Einstein frame 
and the energy density and the pressure of the perfect fluid in the Einstein frame are given by
\begin{align} 
    \rho &= \frac{\A^{1/2}}{(\A - 2\B X)^{1/2}}
        \qty[ \A (\A - \A_{,X}X + 2\B_{,X}X^2) \tilrho + 3 (\A - 2\B X) \A_{,X}X \tilP ] \,,
\\
    P &= A^{3/2} (\A - 2\B X)^{1/2} \tilP \,.
\end{align} 
Solving these relations inversely, we obtain
\begin{align} 
    \tilrho &= \frac{(\A - 2\B X)^{1/2}}{\A^{5/2}(\A - \A_{,X}X + 2\B_{,X}X^2) } (\A \rho - 3 \A_{,X}X P) \,,
\\
    \tilP &= \frac{P}{A^{3/2} (\A - 2\B X)^{1/2}} \,.
\end{align} 
If the perfect fluid is barotropic, then it's equation of state parameter $\tilde{w}=\tilde{P}/\tilde{\rho}$ in Jordan frame and $w=P/\rho$ in the Einstein frame are related as
\begin{equation}
	w = {\frac{( \A - 2\B X ) \tilde{w} }
			  { \A -( 1-3\tilde{w} ) \A_{,X}X +2\B_{,X}X^2 - 6\tilde{w} \A_{,X} \B X^2/\A } } \,. 
\end{equation} 
For a generic conformal transformation $\gtilde_{\mu\nu}=\A(\phi,X)g_{\mu\nu}$, one obtains the relation
\begin{equation}
w=\frac{\tilde{w}}{1 {-} (1-3\tilde{w})\A_{,X}X {/\A}}\,.
\end{equation}
Unless and until $\tilde{w}=1/3$, i.e. the fluid is radiation, $w\neq\tilde{w}$. The fact that for a generic conformal transformation both the frames perceive radiation fluid simultaneously is reflective of the conformal symmetry of the electromagnetic Lagrangian. In particular, notice that for a simple field dependent conformal transformation $\gtilde_{\mu\nu}=\A(\phi)g_{\mu\nu}$, one has the relations
\begin{equation}
\rho=\tilde{\rho}\A^2\,\,,\,\,P=\tilde{P}\A^2\,\,,\,\,w=\tilde{w}\,,
\end{equation}
as expected.

\bibliographystyle{JHEP}
\bibliography{references}

\end{document}